\documentclass[useAMS,usenatbib]{mn2e}
\usepackage{times}
\usepackage{graphicx}
\usepackage{amsfonts} 
\def\ltsima{$\; \buildrel < \over \sim \;$}
\def\lsim{\lower.5ex\hbox{\ltsima}}
\def\gtsima{$\; \buildrel > \over \sim \;$}
\def\gsim{\lower.5ex\hbox{\gtsima}}
\def\ga{\mathrel{\hbox{\rlap{\hbox{\lower4pt\hbox{$\sim$}}}\hbox{$>$}}}}
\def\la{\mathrel{\hbox{\rlap{\hbox{\lower4pt\hbox{$\sim$}}}\hbox{$<$}}}}

\newcommand{\cMpch}{~h^{-1}~\mbox{comoving Mpc}}
\newcommand{\Msunh}{~h^{-1}~\mbox{M}_{\odot}}
\newcommand{\Msun}{~\mbox{M}_{\odot}}
\newcommand{\gadget}{{\sc gadget-2}}
\newcommand{\cmbfast}{{\sc cmbfast}}
\newcommand{\cloudy}{{\sc cloudy}}

\title[Fast Large-Scale Reionization Simulations]
{Fast Large-Scale Reionization Simulations}
\author[Thomas et al.,]{Rajat M.
Thomas\footnote{E-mail:thomas@astro.rug.nl}, 
Saleem Zaroubi\footnote{saleem@astro.rug.nl},
B. Ciardi\footnote{ciardi@MPA-Garching.MPG.DE} 
Andreas H. Pawlik \footnote{pawlik@strw.leidenuniv.nl} \& LOFAR GROUP\\
Kapteyn Astronomical Institute, Landleven 12,Groningen 9747 AD, The
Netherlands \\ 
Max-Planck-Institut fuer Astrophysik,
Karl-Schwarzschild-Strasse 1, D-85748 Garching b. Muenchen, Germany}

\author[Thomas et al., ]{ Rajat M. Thomas$^{1}$\thanks{E-mail:
thomas@astro.rug.nl}, Saleem Zaroubi$^{1}$, Benedetta Ciardi$^{2}$,
Andreas H. Pawlik $^3$, \newauthor Panagiotis Labropoulos$^{1}$, Vibor
Jeli{\'c}$^{1}$, Gianni Bernardi$^{1}$, Michiel A. Brentjens$^{4}$,
\newauthor A.G. de Bruyn$^{1,4}$, Geraint J.A. Harker$^{1}$, Leon V.E.
Koopmans$^{1}$, Garrelt Mellema$^{5}$, \newauthor V.N. Pandey$^{1}$, Joop
Schaye$^{3}$, Sarod Yatawatta$^{1}$\\ $^{1}$Kapteyn Astronomical
Institute, University of Groningen, P.O. Box 800, 9700 AV Groningen,
the Netherlands\\ $^{2}$Max-Planck Institute for Astrophysics,
Karl-Schwarzschild-Stra\ss e 1, 85748 Garching, Germany\\ $^{3}$Leiden
Observatory, Leiden University, PO Box 9513, 2300 RA Leiden, the
Netherlands\\ $^{4}$ASTRON, Postbus 2, 7990 AA Dwingeloo, the
Netherlands\\ $^{5}$Stockholm Observatory, AlbaNova University Center,
Stockholm University, SE-106 91, Stockholm, Sweden }

\begin{document}

\date{}
\pagerange{\pageref{firstpage}--\pageref{lastpage}} \pubyear{2008}

\maketitle

\label{firstpage}

\begin{abstract}
  
 We present an efficient method to generate large simulations of the
 Epoch of Reionization (EoR) without the need for a full 3-dimensional
 radiative transfer code. Large dark-matter-only simulations are
 post-processed to produce maps of the redshifted 21cm emission from
 neutral hydrogen. Dark matter haloes are embedded with sources of
 radiation whose properties are either based on semi-analytical
 prescriptions or derived from hydrodynamical simulations. These
 sources could either be stars or power-law sources with varying
 spectral indices. Assuming spherical symmetry, ionized bubbles are
 created around these sources, whose radial ionized fraction and
 temperature profiles are derived from a catalogue of 1-D radiative
 transfer experiments. In case of overlap of these spheres, photons
 are conserved by redistributing them around the connected ionized
 regions corresponding to the spheres. The efficiency with which these
 maps are created allows us to span the large parameter space
 typically encountered in reionization simulations. We compare our
 results with other, more accurate, 3-D radiative transfer simulations
 and find excellent agreement for the redshifts and the spatial scales
 of interest to upcoming 21cm experiments. We generate a contiguous
 observational cube spanning redshift 6 to 12 and use these
 simulations to study the differences in the reionization histories
 between stars and quasars. Finally, the signal is convolved with the
 LOFAR beam response and its effects are analyzed and
 quantified. Statistics performed on this mock data set shed light on
 possible observational strategies for LOFAR.

\end{abstract}

\begin{keywords} quasars: general -- cosmology: theory -- observation --
diffuse radiation -- radio lines: general.  \end{keywords}

\section{Introduction}

The history of our Universe is largely unknown between the surface of
last scattering ($z\approx1100$) down to a redshift of about
6. Because of the dearth of radiating sources and the fact that we
know very little about this epoch, it is often referred to as the
``dark ages''. Theoretical models suggest that around redshifts 10 --
20, the first sources of radiation appeared that subsequently
reionized the Universe. Two different experiments provide the bounds
for this epoch of reionization (EoR); the high polarization component
at large spatial scales of the temperature-electric field (TE)
cross-polarization mode of the cosmic microwave background (CMB)
providing the upper limit for the redshift at $z \approx 11$
\citep{page07} and the rapid increase in the Lyman-$\alpha$ optical
depth towards redshift 6, observed in the spectrum of high redshift
quasars \citep{fan06}, the lower limit. Although the redshifted 21cm
hyperfine transition of hydrogen was proposed as a probe to study this
epoch decades ago \citep{sz75}, the technological challenges to make
these observations possible are only now being realised. In the
meantime, theoretical understanding of the EoR has improved greatly
\citep{hogan79,scott90,madau97}.  Over the past few years there have
been considerable efforts in simulating the 21cm signal from the Epoch
of Reionization. Almost all of the methods employed in simulating the
21cm involve computer intensive full 3-D radiative transfer
calculations
\citep{otvet,crash,simpX,rsph,ftte,art,zeus,flash,c2ray,zahn07,mesinger07,pawlik08}.

Theories predict that the process of reionization is complex and
sensitively dependent on many not-so-well-known parameters. Although
 stars may be the most favoured of reionization
sources, the role of mini-quasars (miniqsos), with the central black
hole mass less than a few million solar masses, are debated
\citep{nusser05,zaroubi05,kuhlen05,rajat08}.  Even if the nature of
the sources of radiation would be relatively well constrained, there are a
number of ``tunable'' parameters like the photon escape fraction,
masses of these first sources, and so on, that are not well
constrained. 

In a couple of years, next generation radio telescopes like
LOFAR\footnote{www.lofar.org}\footnote{www.astro.rug.nl/\~~LofarEoR} and
MWA\footnote{http://www.haystack.mit.edu/ast/arrays/mwa/} will be
tuned to detect the 21cm radiation from the EoR. Although the designs
of these telescopes are unprecedented, the prospects for successfully
detecting and mapping neutral hydrogen at the EoR critically depends
on our understanding of the behaviour and response of the instrument,
the effect of diffuse polarized Galactic \& extra-galactic emission,
point source contamination, ionospheric scintillations, radio
frequency interference (RFI) and, not least, the characteristics of
the desired signal. A good knowhow of the above phenomena would enable
us to develop advanced signal processing/extraction algorithms, that
can be efficiently and reliably implemented to extract the signal.  In
order to test and confirm the stability and reliability of these
algorithms, it is imperative that we simulate, along with all the
effects mentioned above, a large range of reionization scenarios.

\begin{figure}
\centering
\hspace{0cm}
\includegraphics[width=.5\textwidth]{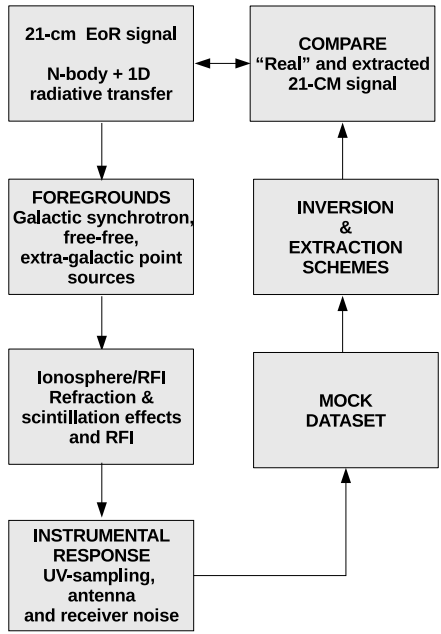}
\vspace{-3cm}
\caption{The big picture: This simple flow diagram encapsulates the
essence of the ``LOFAR EoR simulation pipeline''. Starting from the
generation of the the cosmological EoR signal, the pipeline includes
the addition of foreground contaminants like Galactic synchrotron and
free-free radiation and other point sources, and the LOFAR antenna
response. This ``mock'' data set is then used to extract the signal
using various inversion algorithms and the result is then compared to
the original ``uncorrupted'' signal in order to study the accuracy and
stability of the inversion schemes employed. }
\label{fig:bigpic}
\end{figure}

Fig.~\ref{fig:bigpic} shows the basic building blocks of the
simulation pipeline being built for the LOFAR-EoR experiment. This
paper basically constitutes the first block, i.e., simulation of the
cosmological 21cm EoR signal. This then passes through a sequence of
blocks like the foreground simulation \citep{jelic08}, instrument
response and extraction (Lambropoulous et al., \emph{in prep}). The
extracted signal is then compared with the original signal to quantify
the performance of the extraction scheme. This process needs to be
repeated for various reionization scenarios to avoid any bias the
extraction scheme would have if only a subsample of all possible
signal characteristics were used.

Simulating observing windows as large as the Field of View (FoV) of
LOFAR ($\sim 5^{\circ} \times 5^{\circ} $) and for frequencies
corresponding to redshift 6 to 12 is a daunting task for conventional
\hbox{3-D} radiative transfer codes because of multiple reasons such
as a requirement for high dynamic range in mass for the sources of
reionization, their large number towards the end of reionization and
the size of the box which strains the memory of even the largest
computer cluster. In order to facilitate the simulation of such large
mock data sets for diverse reionization scenarios, we need to
implement an approximation to these radiative transfer methods that
mimic the ``standard'' \hbox{3-D} simulations to good accuracy.  It
was clear from the onset that the details of the ionization fronts
like its complex non-spherical nature will not be reproduced by the
semi-analytical approach that we propose here. But the argument
towards overlooking this discrepancy is that when the outputs of our
semi-analytical approach and that of a \hbox{3-D} radiative transfer
code are passed through the machinery of the LOFAR-EoR pipeline, they
are experimentally indistinguishable. The reason being the filtering
nature of the telescope's point spread function (PSF) across the sky
and the substantial bandwidth averaging along the frequency/redshift
direction that is needed to recover the signal, smoothes out the
structural details captured by \hbox{3-D} codes.

Recently, several authors \citep{zahn07,mesinger07} have proposed
schemes to reduce the computational burden of generating relatively
accurate 21cm maps.  These methods do fairly well, although there are
some caveats, like for example the intergalactic medium (IGM)
ionization being treated as binary, i.e., the IGM is either ionized or
neutral \citep{mesinger07}. Although this might be the case for
stellar-like sources, others with a power-law component could exhibit
an effect on the IGM wherein the ionizing front is extended and hence
this assumption need not hold \citep{zaroubi05,rajat08}. Added to
this, the schemes presented in order to compute the 21cm maps, make
the assumption that the spin temperature of hydrogen, $T_s$ is much
larger than the CMB temperature. Towards the end of reionization ($z <
8 $) this might very well be valid. But the dawn of reionization would
see a complex spatial correlation of IGM temperatures with the sources
of radiation, its clustering and spectral energy distributions
\citep{venky01,zaroubi07,rajat08,pritchard07}. In the current paper we
have assumed that the spin temperature is coupled to the kinetic
temperature and that they are much higher than the CMB temperature. At
higher redshifts this need not be a valid assumption. The effects of
heating by different types of radiative sources on the IGM and the
coupling (both Ly-$\alpha$ and collisional) between the spin and
kinetic temperature will be simulated accurately in a follow-up paper
\emph{Thomas et al., in prep}, using the same scheme, but now applied
to heating.

In this paper we propose a method of post processing numerical
simulations in order to rapidly generate realistic 21cm maps. Briefly,
the algorithm consists of simulating the ionization fronts created by
the ``first'' radiative sources for a range of parameters which
include the power spectrum, source mass function and clustering. We
then identify haloes in the outputs of N-body simulations and convert
them to a photon count using semi-analytical prescriptions, or using
the photon count derived from a Smoothed Particle Hydrodynamics (SPH)
simulation.  Depending on the photon count and the spectrum, we embed
a sphere around the centre-of-mass (CoM) of the halo whose radial
profile matches that of a profile from the table created by the 1-D
radiative transfer code of \citet{rajat08}. Appropriate operations are
carried out to conserve photon number. Since, the basic idea is to
expand bubbles around locations of the sources of radiation, we call
this method BEARS (Bubble Expansion Around Radiative Sources). For the
sake of brevity and comparison with full \hbox{3-D} radiative transfer
codes, we restrict ourselves to monochromatic radiative transfer with
a fixed temperature.  A following paper will include a full spectrum
along with the temperature evolution. The results of our semi-analytic
scheme will be compared to those obtained with the full \hbox{3-D}
Monte Carlo radiative transfer code CRASH \citep{crash,maselli03}.

In \S\ref{sec:simulations} we describe the various steps involved in
implementing the BEARS algorithm on the outputs of N-body
simulations. We describe the specifications of the N-body simulations,
the 1-D radiative transfer code used to produce the catalogue of
ionization profiles, the algorithm employed to embed the sources and
finally an illustrative example of the procedure to correct for the
overlap of ionized bubbles. In \S\ref{sec:comparison} the fully
\hbox{3-D} cosmological radiative transfer code CRASH is summarized
and the results of its qualitative and statistical comparison with
BEARS are discussed.  \S\ref{sec:datacube} describes the method of
generating the cube with maps of the brightness temperature ($\delta
T_b$) at all the frequencies that will be observed by an EoR
experiment.  In \S\ref{sec:starsVquasars} we use the simulation of the
cube, with maps of the sky at different frequencies, to study the
difference between two popular sources of reionization, i.e., stars
and quasars.  These maps in the cube are then filtered through the
LOFAR antenna response to output the final data cube in
\S\ref{sec:instrumental}. Finally in \S\ref{sec:conclusions} we
summarize our results and outline further improvements that need to be
made in our approach in order to start exploring the large parameter
space involved in reionization studies.

\section{Simulations: The BEARS algorithm}
\label{sec:simulations}
In this section the various components of the simulation that lead
towards 21cm brightness temperature ($\delta T_b$) maps are
explained. The very first step of the entire process consists of
generating a catalogue of 1-D ionization profiles for different
masses/luminosities of the source, their spectra and the density
profiles that surround them at different redshifts. Although we assume
the density around each source to be constant, we do vary its value as
detailed below. Given the locations of the centres of mass of haloes,
the photon rate emanating from that region is calculated based on a
semi-analytical description (discussed in section
\ref{sec:starsVquasars}). Given the spectrum, luminosity and the
overdensity around the source, a spherical bubble is embedded around
that pixel whose radial profile is selected from the table. The
justification for this simple-minded approach is that the Universe is
relatively homogeneous in density at these high redshifts on the
scales probed by upcoming EoR experiments. Therefore, we can assume
spherical symmetry in our construction of the ionized regions.

In the following subsections we summarize the N-body simulations employed and
the 1-D radiative transfer code that was used to generate the
catalogue, and we describe in detail the algorithm used to embed the
spherical bubbles and the method used to conserve photons in case of
overlap.

\subsection{N-body/SPH simulations}
\label{sec:nbody}
We used a modified version of the N-body/TreePM/SPH code \gadget\
(\citealp{Springel:2005}) to perform a dark matter (DM) cosmological
simulation containing $512^3$ particles in a box of size $100 \cMpch$
and a DM+SPH cosmological simulation containing $256^3$ DM and $256^3$
gas particles in a box of size $12.5\cMpch$. The DM particle masses
were $4.9\times 10^8 \Msunh$ and $6.3 \times 10^6 \Msunh$,
respectively.  \par Initial particle positions and velocities were
obtained from glass-like initial conditions using \cmbfast\ (version
4.1; \citealp{Seljak:1996}) and employing the Zeldovich approximation
to linearly evolve the particles down to redshift z = 127. We assumed
a flat $\Lambda$CDM universe and employed the set of cosmological
parameters $\Omega_m = 0.238$, $\Omega_b = 0.0418$, $\Omega_\Lambda =
0.762$, $\sigma_8 = 0.74$, $n_s=0.951$ and $h=0.73$, in agreement with
the WMAP 3-year observations (\citealp{Spergel:2007}).  Data was
generated at $50$ equally spaced redshifts between $z = 20$ and $z =
6$.  Halos were identified using the Friends-of-Friends algorithm
(\citealp{Davis:1985}), with linking length $b = 0.2$.

\par The gas in the DM+SPH simulation is of primordial composition,
with a hydrogen mass fraction $X = 0.752$ and a helium mass fraction
$Y = 1-X$.  Radiative cooling and heating are included assuming
ionisation equilibrium, using tables\footnote{For a detailed
description of the implementation of radiative cooling see
\cite{Wiersma:2008}.} generated with the publicly available package
\cloudy\ (version 05.07 of the code last described by
\citealp{Ferland:1998}).  The gas is allowed to cool by collisional
ionisation and excitation, emission of free-free and recombination
radiation and Compton cooling off the cosmic microwave
background. Molecular cooling (by hydrogen and deuterium) is prevented
by the inclusion of a soft (i.e.\ cut off at the Lyman-limit) UV
background. We employed the star formation recipe of
\cite{Schaye:2008}, using a \cite{Chabrier:2003} initial mass function
(IMF) with mass range $[0.1, 100]\Msun$.

\par For further analysis, SPH particle masses were assigned to
uniform meshes of size $64^3, 128^3$ and $256^3$ cells using TSC
(\citealp{Hockney:1988}) and the gas densities were calculated. The
density field was smoothed on the mesh with a Gaussian kernel with
standard deviation $\sigma_G =12.5 / 512$ comoving ${\rm Mpc}/h$. Each
cell was further assigned a hydrogen-ionizing luminosity according to
the stellar mass it contained. Star particles were treated as simple
stellar populations and their luminosity was calculated with the
population synthesis code of \cite{Bruzual:2003}. We used stellar
masses and ages as determined by the simulation, a Chabrier IMF
consistent with the star formation recipe and assumed a fiducial metal
mass fraction $Z = 0.0004$.

\subsection{1-D radiative transfer (RT) code.}

The catalogue of ionization profiles for different redshifts (which
also translates to different densities at a given redshift), spectral 
energy distributions (SED), times of evolution and masses/luminosities, 
was created using the 1-D radiative transfer code developed by
\citet{rajat08}. 

Following \cite{fuku94}, a set of rate equations are solved at every
cell as a function of time. The equations follow the time-evolution of
H$_\mathrm{I}$, H$_\mathrm{II}$, He$_\mathrm{I}$, He$_\mathrm{II}$,
He$_\mathrm{III}$ and temperature at every grid cell. The ionization
rates are integrals over the spectrum and the cross sections
of the various species. These were pre-computed and stored in a table
to facilitate faster execution. Case-B recombination coefficients are 
used to calculate the rate at which hydrogen recombines.

The 1-D radiative transfer code starts the simulation at $\rmn
R_{start}$, typically $0.1$~physical $\rmn{kpc}$ from the location of
the source. All hydrogen and helium is assumed to be completely
ionized inside this radius, $\rmn R_{start}$. Each cell is then
updated for time $\Delta t$. This $\Delta t$ is not the intrinsic
time-step used to solve the differential equation itself because that
is adaptive in nature and varies according to the tolerance limit set
in the ordinary differential equation (ODE) solver.  On the other
hand, the $\Delta t$ here decides for how long a particular cell
should evolve before moving on to the next. The code is causal in the
sense that cell $i$+1 is updated after cell $i$. The light travel time
is not taken into consideration explicitly since the ionization front
(I-front) is typically very subluminal. Hence, all cells are updated
to time $n \Delta t$ at the $n^{th}$ time-step.

After all cells except the last cell $i_{Rmax}$ have been updated to
time $\Delta t$, the resulting values are stored and then passed on as
initial conditions for the evolution of the cell in the next interval
of time. The update of all $n_{cell}$ cells is repeated ~$n$~ times
such that ~$n\Delta t = t_\rmn{source}$, where $t_\rmn{source}$ is the
life time of the radiating source. The various quantities of interest
can be stored in a file at intervals of choice.

A radial coverage of $R_{max}$ is chosen \textit{a priori} which,
depending on the problem, can be set to any value. Typically we do not
need to solve the radiative transfer beyond ten comoving
mega-parsecs. We use an equally spaced grid in radius with a
resolution of $\Delta r$  which, like the time resolution, is decreased
to half its value until it meets a given convergence criterion, which
here is that the final position of the I-front converges to within
$0.5\%$.

The implementation of the code is modular. Hence it is straightforward
to include different spectra corresponding to different ionizing sources.
Our 1-D code can handle X-ray photons and the secondary
ionization and heating it causes and therefore performs well for both
high (quasars) and low energy (stars) photons. Further details of the
code can be found in \citet{rajat08}.

\subsection{Embedding the 1-D radiative code into the simulation box}
\label{sec:bears}
In the following section we discuss the algorithm employed to expand
reionization bubbles around the locations of radiative sources. The
numerical simulation provides us with the density field at
every grid point, the centres of mass of the haloes identified by the
FoF algorithm, the velocities of these particles and, if the simulation
also contains gas, the ionizing luminosity associated with the halo.

Equipped with this information about the simulation box at a given redshift,
we follow the steps enumerated below to embed the ionized bubbles around
the source locations:

\begin{enumerate}
\item Given the redshift, ionizing luminosity and the time for which the
ionization front should evolve (this depends on the type of source), select a
corresponding file from the catalogue of ionization profiles generated
earlier.

\item The sources are usually in an overdense region and the density
around the source follows a profile. Since the profiles vary from
source to source we use the following approximation: we calculate the
overdensity around the source for a radius $R_\rmn{od}$ (where the
subscript ``od'' stands for overdensity). We then assume that the
source is embedded in a uniform density whose value is the average
overdensity within the radius $R_{od}$. This naturally translates in
selecting the same ionization profile but now from the table at a
higher redshift. This radius is estimated as described in
\S\ref{sec:avgrad}.

\item At lower redshifts there is considerable overlap between
bubbles. Thus, in order to conserve the number of photons, we need to
redistribute the photons that ionize the overlapped regions to other
regions which are still neutral. The details of this correction
process are given below (see \S\ref{sec:corr_overlap}).

\item When computing the reionization history, ionized regions are
mapped one-to-one from the current simulation snapshot onto the next.
Note that ionization due to recombination radiation has not been
included.

\end{enumerate}

\subsubsection{Estimating the averaging radius}
\label{sec:avgrad}

The radius $R_\rmn{od}$ is calculated as follows: 
\begin{itemize}
\item For any given source (we choose the largest) in the
  box we perform the radiative transfer and compute the ionization
  profile using the (exact) spherically averaged radial density
  profile of the surrounding gas;
\item The density within a radius $R_\rmn{od}$, which we initially
  choose to be very small, is spherically averaged around the source
  and the ionization profile corresponding to this mean density is
  selected from the catalogue. This step is repeated for increasing
  values of $R_\rmn{od}$ until the extent of the ionization profile
  matches that of the ``exact'' radiative transfer calculation done in
  the previous step. We will refer to the resulting radius as the
  calibrated $R_\rmn{od}(cal)$;
\item $R_\rmn{od}$ for all the other sources are calculated by scaling
  them with the luminosity of the source according to $R_\rmn{od} =
  R_\rmn{od}(\rmn{cal}) \left ( \frac{L_\rmn{source}}{L_\rmn{cal}}
  \right )^{1/3}$, where $L_\rmn{source}$ and $L_\rmn{cal}$ are the
  luminosities of the source under consideration and the calibration
  source respectively.
\end{itemize}
A comment to be made here is that because the dynamic range in the
halo masses derived from the simulation is not very high the value of
$R_\rmn{od}$ does not vary considerably between sources.

\subsubsection{Correction for overlap}
\label{sec:corr_overlap}
\begin{figure}
\centering
\hspace{0cm}
\includegraphics[width=.5\textwidth]{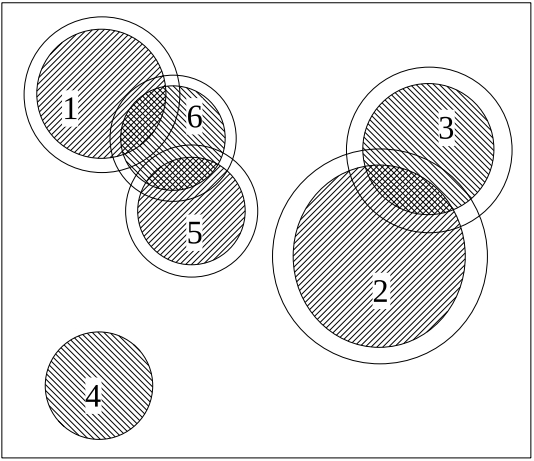}
\vspace{.1cm}
\caption{A cartoon depicting the distribution and overlap of bubbles
in the simulation. The number within each circle is just an associated
identity of the circle. In the above figure circles (1:6:5), (2:3) and
(4) form three different ``connected'' regions. Each circle of the
region is expanded just enough that the size of the region expanded
corresponds to the area of overlap. In this manner we hard-wire the
conservation of photons into the algorithm.}
\label{fig:incidence}
\end{figure}

At lower redshifts the sources become more massive and more
numerous. This causes considerable overlap between spheres of close-by
sources. The regions of overlap correspond to some number of
unused photons (depending on the density). The relative simplicity of
the implementation of the algorithm arises due to the fact that the
ionization fronts created by most sources are ``sharp'', at least for
the resolutions we are able to simulate. If the ionization front is
extended, we can modify the procedure below to accommodate it. The
procedure for correction is illustrated with the help of a simple
example shown in Fig.~\ref{fig:incidence}:
\begin{itemize}
\item Create an analogue to an incidence/admittance matrix \textbf{A}
(ref equation \ref{eq:incidence} for the connection between ionized regions of
sources): The matrix \textbf{A} is a square matrix
of size $N_\rmn{source} \times N_\rmn{source}$, whose elements are set to
1 if two sources overlap and 0 otherwise. Here $N_\rmn{source}$ is the
total number of sources in the box.

\begin{equation}
 \bf{A} = \left( \begin{array}{*{10}c}
     1 & 0 & 0 & 0 & 1 & 1  \\
     0 & 1 & 1 & 0 & 0 & 0  \\
     0 & 1 & 1 & 0 & 0 & 0  \\
     0 & 0 & 0 & 1 & 0 & 0  \\
     1 & 0 & 0 & 0 & 1 & 1  \\
     1 & 0 & 0 & 0 & 1 & 1  \\
\end{array} \right)
\label{eq:incidence}
\end{equation}

\item Segment the simulation box into regions that are connected. This
is done by identifying rows that are identical in the matrix
\textbf{A}. Thus, for the example shown in Fig.~\ref{fig:incidence}
there are three different ``connected regions'', Reg1:(1,6,5), Reg2:(2,3)
and Reg3:(4).

\item Calculate the total volume of the overlap zones in each of the
connected regions. For example for Reg1, we have
Overlap(1,6)+Overlap(5,6).

\item The sizes of all the bubbles in a particular connected region is 
increased so as to entail a volume that corresponds to the average overlapped
region, i.e., total overlapped region divided by the number of bubbles in that
connected region. 

\item The sizes of the bubbles are iteratively increased until the
volume of the regions that are ``newly ionized'', i.e., excluding the
regions ionized by the sources before the correction, equals the total
overlapped volume. This ensures a homogeneous redistribution of unused
photons all around the region. The unshaded bubbles in Reg1:(1,6,5)
and Reg2:(2,3) in Fig.~\ref{fig:incidence} depict the expansion of the
bubbles.

\end{itemize}

\section{Comparison with a full \hbox{3-D} Radiative Transfer code.}
\label{sec:comparison}

The BEARS algorithm is used to generate a series of cubes of ionized
fractions at different redshifts and resolutions for the case in which
the radiative sources are stars. In this section these cubes are
compared with those obtained from the full \hbox{3-D} radiative
transfer code CRASH using the same source list i.e., source locations,
luminosities and spectra. In all comparisons made below we make use of
ionizing luminosities derived from the N-Body/SPH simulation (refer
\S\ref{sec:nbody}). We have assumed the spectra to be monochromatic at
13.6~$\mathrm{eV}$. We first summarize the essentials of the
\hbox{3-D} radiative transfer code CRASH and then discuss a set of
visual and statistical comparisons between the results generated using
these two approaches.

All comparisons except for the case of $12.5\cMpch$, $128^3$ box, in
this section is done on boxes where we have not included the
reionization history . One of the major reasons for this is that a
full 3-D radiative transfer code like CRASH would require enormous
computational resources to achieve this task. And on the other hand,
it is easier to judge the performance of BEARS when there a large
number of sources overlapping at lower redshifts, whereas if the
reionization history is included, the lower redshifts would be
completely ionized and most of the interesting characteristics of the
comparison would be wiped out.

\subsection{CRASH: An overview}

CRASH is a \hbox{3-D} ray-tracing radiative transfer code based on Monte Carlo
(MC) techniques that are used to sample the probability distribution
functions (PDFs) of several quantities involved in the calculation,
e.g.\ spectrum of the sources, emission direction and optical depth. The
MC approach and the code architecture enables applicability over a
wide range of astrophysical problems and allows for additional physics to
be incorporated with minimum effort.  The propagation of the
ionizing radiation can be followed through any given H/He static
density field sampled on a uniform mesh. At every grid point and time
step the algorithm computes the
variations in temperature and ionization state of the gas. The code
allows for the possibility of the addition of multiple point sources
at any specified point in the box, and also diffuse radiation
(e.g.\ the ultraviolet background or the radiation produced by H/He
recombinations) can be self-consistently incorporated.

The energy emitted by point sources in ionizing radiation is
discretized into photon packets, beams of ionizing photons, emitted at
regularly spaced time intervals. More specifically, the total energy
radiated by a single source of luminosity $L_s$, during the total
simulation time, $t_\rmn{sim}$, is $E_s = \int_0^{t_\rmn{sim}}
L_s(t_s) \rmn{d}t_s$. For each source, $E_s$ is distributed in $N_p$
photon packets, emitted at the source location at regularly spaced
time intervals, $dt = t_\rmn{sim}/N_p$. The time resolution of a given
run is thus fixed by $N_p$ and the time evolution is marked by the
packets' emission: the j-th packet is emitted at time
$t^j_\rmn{em,c}=j\times dt$, with $j=0, ..., (N_p-1)$. Thus, the total
number of emissions of continuum photon packets is $N_\rmn{em,c} =
N_p$. The emission direction of each photon packet is assigned by MC
sampling the angular PDF characteristic of the source. The propagation
of the packet through the given density field is then followed and the
impact of radiation-matter interaction on the gas properties is
computed on the fly. Each time the packet pierces a cell i, the cell
optical depth for ionizing continuum radiation,$\tau^i_c$ , is
estimated summing up the contribution of the different absorbers
($\rmn{H_I, He_I, He_{II}}$). The probability for a single photon
to be absorbed in the $i$th cell is:
\begin{equation}
P(\tau^i_c)=1-e^{-\tau^i_c}.
\end{equation}
The trajectory of the packet is followed until its photon content is
extinguished or, if open boundary conditions are assumed, until it
exits the simulation volume.  The time evolution of the gas physical
properties (ionization fractions and temperature) is computed solving
in each cell the appropriate discretized differential equations each
time the cell is crossed by a packet.The reader is referred to
\citet{maselli03,crash} for details of the code.

\subsection{``Visual'' comparison}

\begin{figure*}
\centering
\hspace{0cm}
\includegraphics[width=.4\textwidth,height=.4\textwidth]{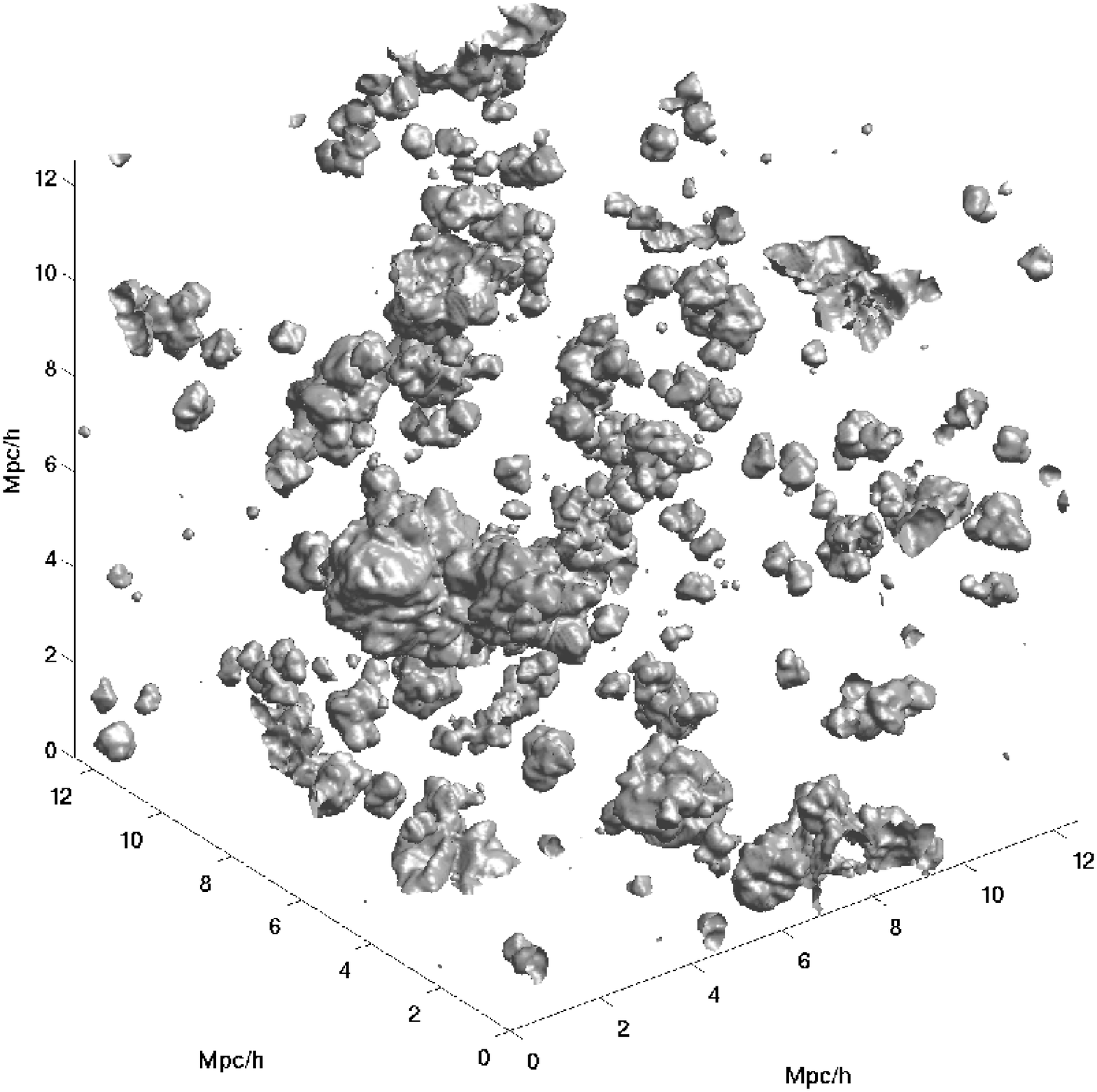}
\hspace{.2cm}
\includegraphics[width=.4\textwidth,height=.4\textwidth]{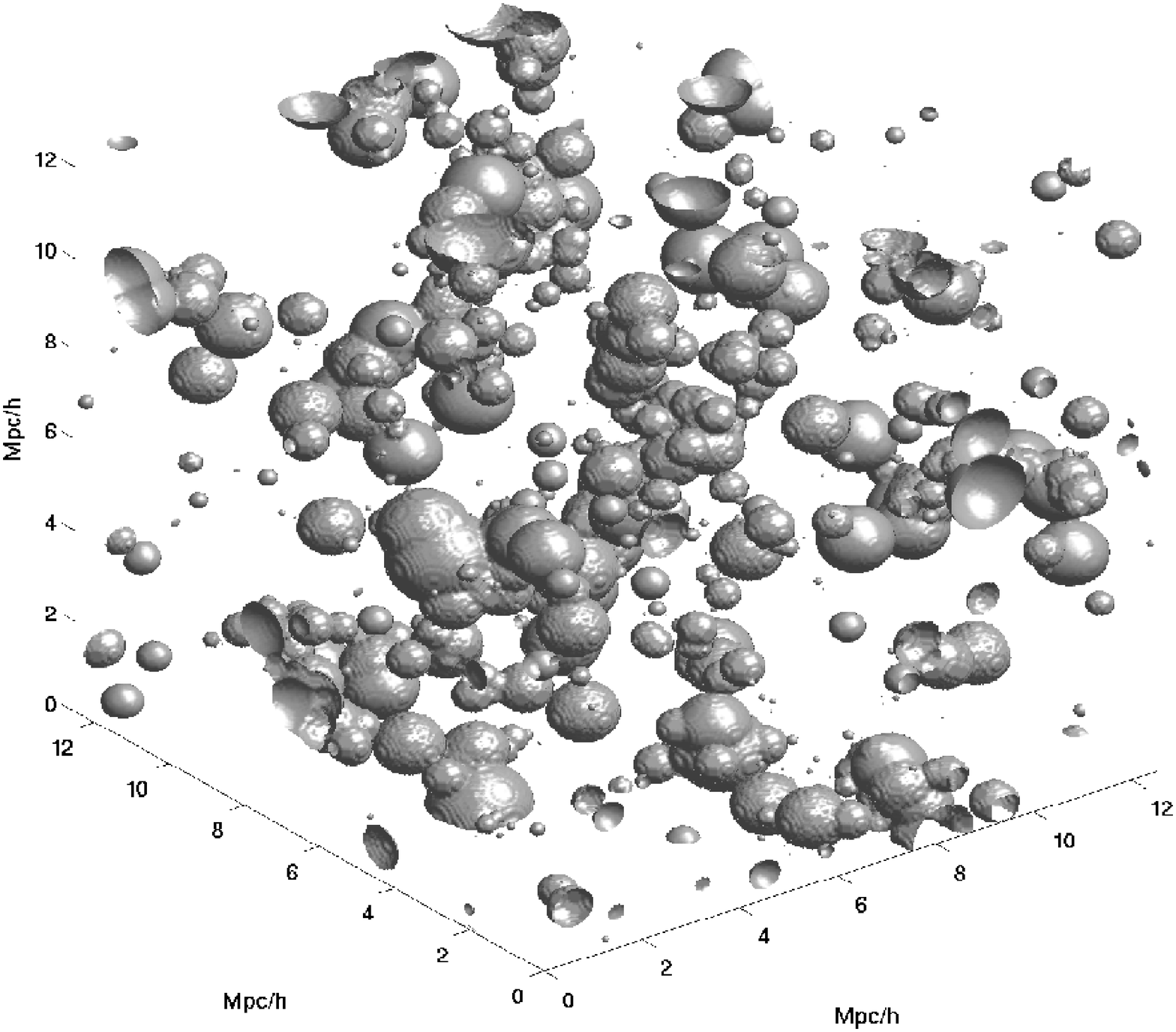}
\vspace{.2cm}
\caption{A \hbox{3-D} visualization of the ionized box from
CRASH (left) and BEARS (right) at redshift z $\approx$ 6 for a boxsize
of $12.5 \cMpch$ at a resolution of $256^3$. As we see, although this
is the redshift for which we expect a large discrepancy, the figures seem
to be morphologically comparable.}
\label{fig:comparered6}
\end{figure*}

Figure~\ref{fig:comparered6} shows a \hbox{3-D} view of the ionized
fraction of hydrogen calculated using CRASH (left panel) and BEARS
(right panel), with isosurfaces shaded dark. The box shown here is a
$256^3$-simulation at a redshift of six for a box with a comoving
length of $12.5 \cMpch$ on each side. Globally the two boxes do look
very similar. Statistics on these cases and others are presented
later. Also bear in mind that this case, i.e., redshift six, is the
one for which we should expect maximum discrepancy between the two
methods. Reasons being: one, the universe is much less homogeneous at
redshift six than at higher redshifts which is contrary to the basic
assumption in BEARS that the IGM is predominantly uniform, two, the
bubbles from these ionizing sources become larger (because the sources
grow more massive and because the gas density decreases) and they
invariably overlap with several others which in our case is dealt with
in an approximate manner as explained before.

\begin{figure}
\hspace{-0.4cm}
\includegraphics[width=.5\textwidth]{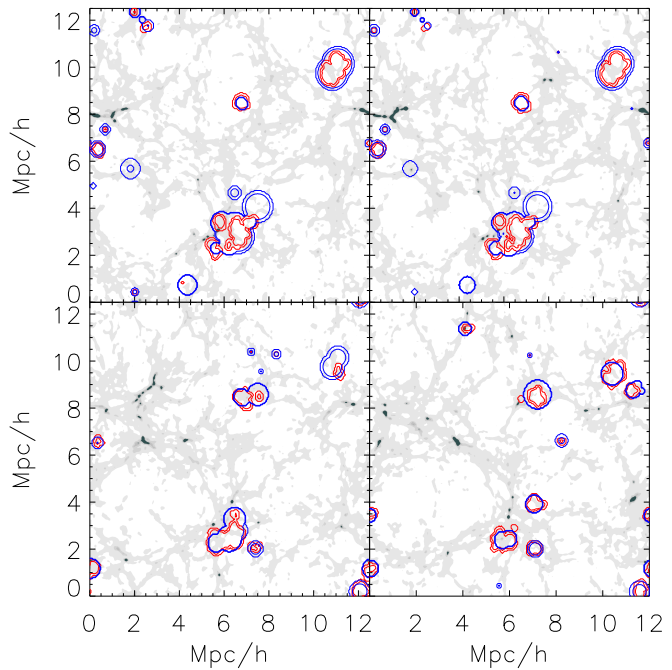}
\vspace{.2cm}
\caption{Four slices (thickness $\approx 0.05 \cMpch$), randomly
selected along a direction in the $12.5 \cMpch$, $256^3$ box, are
plotted that displays the contours (three levels [0, 0.5, 1]) of the
neutral fraction of CRASH (red) and BEARS (cyan) at $z\approx 6$. The
underlying ``light gray'' contours represent the dark matter
overdensities.}
\label{fig:overlaps256Red6}
\end{figure}

\begin{figure}
\centering
\hspace{0cm}
\includegraphics[width=.5\textwidth]{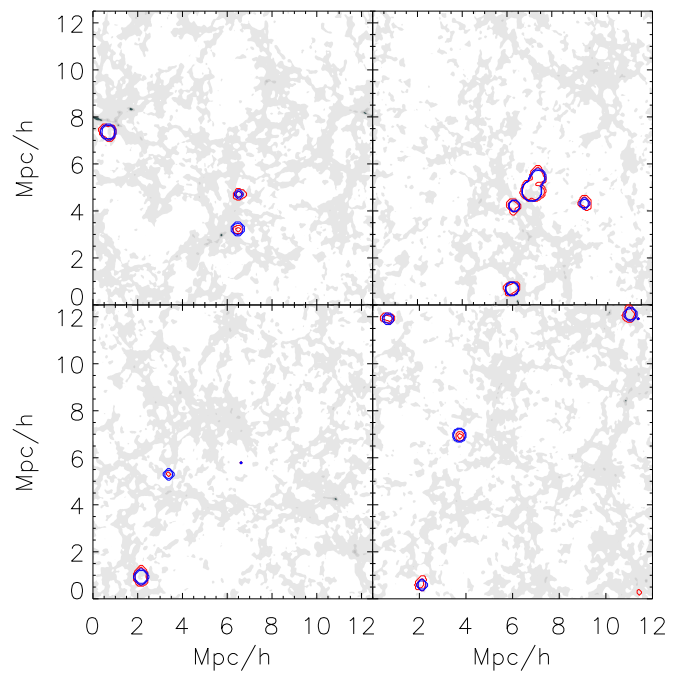}
\vspace{.2cm}
\caption{Same as Fig.~\ref{fig:overlaps256Red6} but for redshift 
$z\approx 9$.}
\label{fig:overlaps256Red9}
\end{figure}

In Figs.~\ref{fig:overlaps256Red6} \& \ref{fig:overlaps256Red9} we
compare the ionization isocontours from BEARS (blue contours) and
CRASH (red contours) for redshifts 6 and 9, respectively. The images
refer to the simulations in a $256^3$ box with side of length
$12.5~\cMpch$. As expected, redshift 6 is structurally the most
complex with many more sources and multiple overlaps. In all these
images we see a distinct feature of the BEARS algorithm, i.e., that
the ionized regions are perfectly spherical when there is no
overlap. Also, in case of overlap, the shapes of ionized regions are
much more regular than those of CRASH. The reason for this discrepancy
is the local inhomogeneity of the underlying density field. As
explained in \S\ref{sec:bears}, the BEARS algorithm averages over a
radius of $R_{od}$ and uses the same density in all directions,
whereas CRASH follows the local density separately in each
direction. With higher resolution the agreement between the two
simulations is expected to decrease because the density field is not
as isotropic, and in situations like this the \hbox{3-D} codes are
better at following the non-spherical nature of the ionizing front.

\begin{figure}
\hspace{-0.4cm}
\includegraphics[width=.5\textwidth]{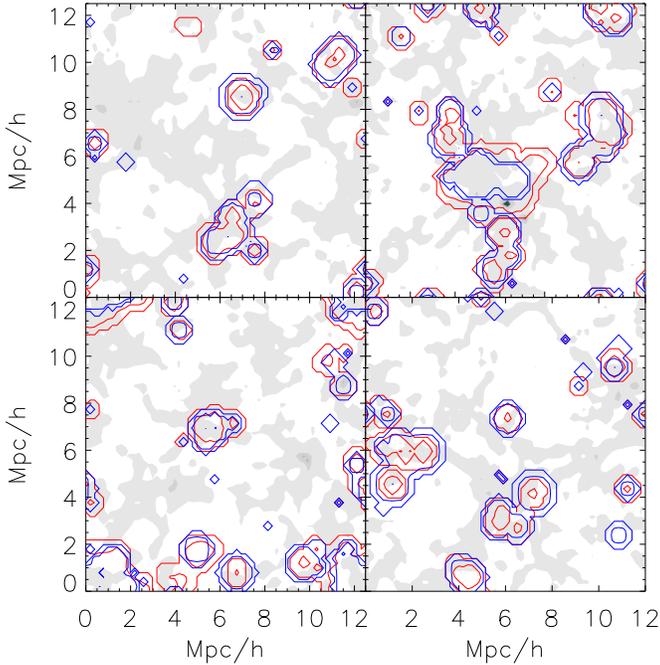}
\vspace{.2cm}
\caption{Four slices (thickness $\approx 0.2 \cMpch$), randomly
selected along a direction in the $12.5 \cMpch$, $64^3$ box, are
plotted that displays the contours (three levels [0, 0.5, 1]) of the
neutral fraction of CRASH (red) and BEARS (cyan) at $z\approx 6$. The
underlying ``light gray'' contours represent the dark matter
overdensities. }
\label{fig:overlapsRed6}
\end{figure}

\begin{figure}
\centering
\hspace{0cm}
\includegraphics[width=.5\textwidth]{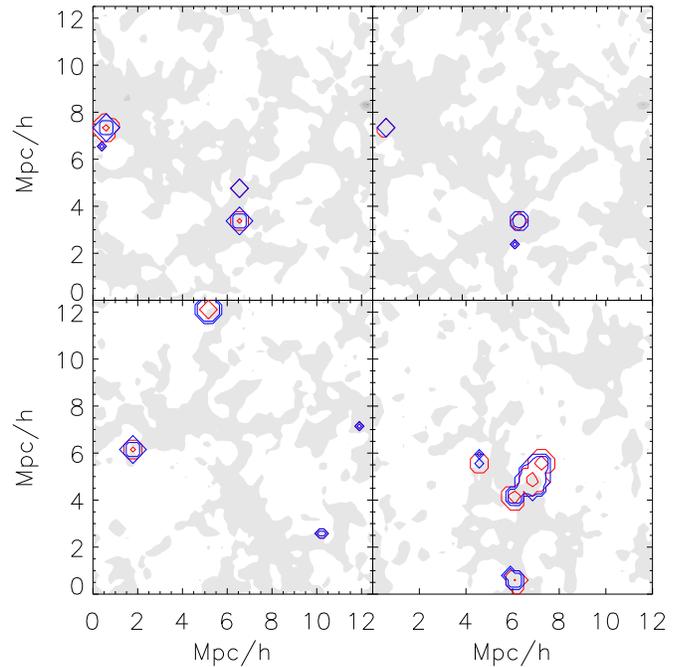}
\vspace{.2cm}
\caption{Same as Fig.~\ref{fig:overlapsRed6} but for redshift
$z\approx 9$.}
\label{fig:overlapsRed9}
\end{figure}

In Figs.\ref{fig:overlapsRed6} and \ref{fig:overlapsRed9}, we compare
BEARS and CRASH for a lower resolution simulation, i.e., $64^3$ in a 
$12.5 \cMpch$ box for redshifts 6 to 9. These figures indeed
show a much better agreement because the detailed structure of the
ionization front traced by CRASH at a higher resolution has been
smoothed out.

\begin{figure}
\hspace{-0.4cm}
\includegraphics[width=.5\textwidth]{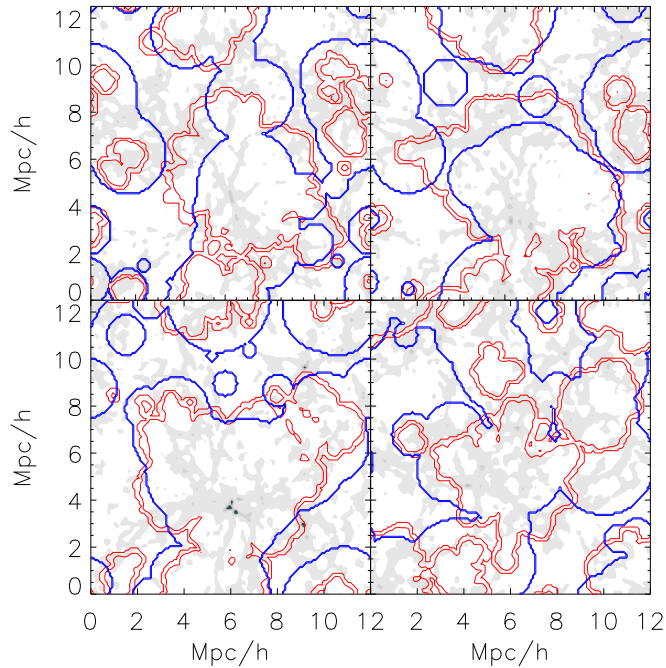}
\vspace{.2cm}
\caption{Four slices (thickness $\approx 0.1 \cMpch$) from a
$12.5\cMpch$, $128^3$ box are randomly selected and contours (three
levels [0, 0.5, 1]) of the neutral fraction of CRASH (red) and BEARS
(blue) at $z\approx 6$ are plotted. The underlying ``light gray''
contours represent the dark matter overdensities. In this figure we
have taken into account the ionization history, i.e., the radiative
transfer is performed on the box which had been partially ionized at
earlier redshifts.}

\label{fig:overlap6withmem}
\end{figure}

\begin{figure}

\hspace{0cm}
\includegraphics[width=.5\textwidth]{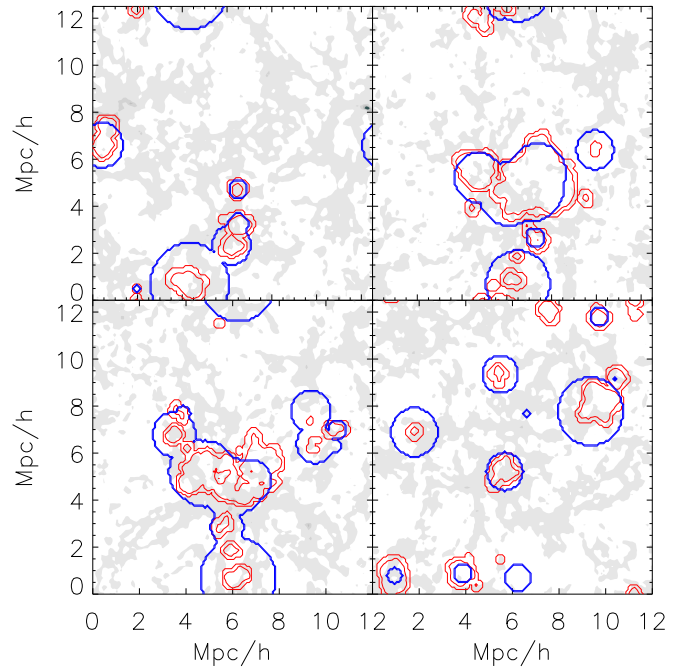}

\hspace{-0.8cm}
\includegraphics[width=.56\textwidth]{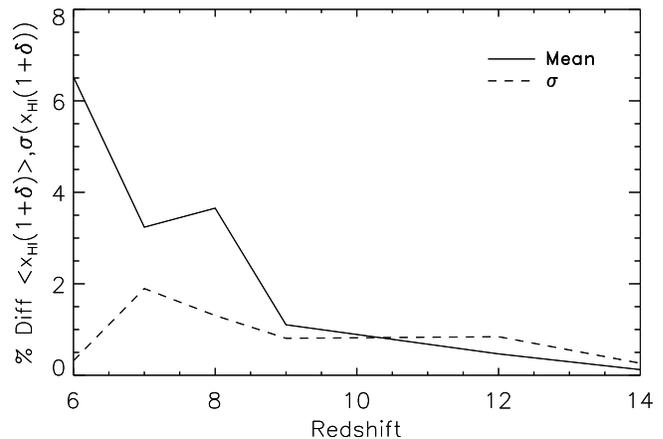}
\caption{ The top panel is same as Fig.~\ref{fig:overlap6withmem} but
for redshift $z\approx 9$. The bottom panel shows the percentage
difference between CRASH and BEARS in the mean (solid line) and
variance (dashed line) of the mass-weighted ionized fraction (ionized
fraction $>$ 0.95) in the $128^3$ box as a function of redshift when
the history of reionization is included.}
\label{fig:overlap9withmem}
\end{figure}

All the previous figures of comparison of CRASH and BEARS were
performed without taking into account the history of
reionization, i.e., radiative transfer was performed on each snapshot
assuming that it had not been previously ionized. In
Figure~\ref{fig:overlap6withmem} and the top panel of
Figure~\ref{fig:overlap9withmem} we plot four slices of the $12.5
\cMpch$, $128^3$ box at redshifts of 6.2 and 9 respectively, which do
include the memory of ionization from previous redshifts. We observe
that the agreement in this case is not as good as in the case
without the history of reionization, but given that this comparison is
made at the lowest redshift of interest and for a high resolution
($12.5 \cMpch$) simulation, the results are acceptable. The bottom
panel of Figure~\ref{fig:overlap6withmem} shows the mean (solid line)
and variance (dashed line) of the mass-weighted ionized fraction as a
function of redshift for a $12.5\cMpch$, $128^3$ box. The reason why the
statistics of mass-weighted ionized fraction agree so well whereas
the contour plots have some descrepancy is that when the underlying
density is high, BEARS estimates the extent of the ionization correctly,
whereas in case of under-dense regions, BEARS overestimates the size
of the bubble.

\begin{figure}
\centering
\hspace{-1cm}
\includegraphics[width=.5\textwidth]{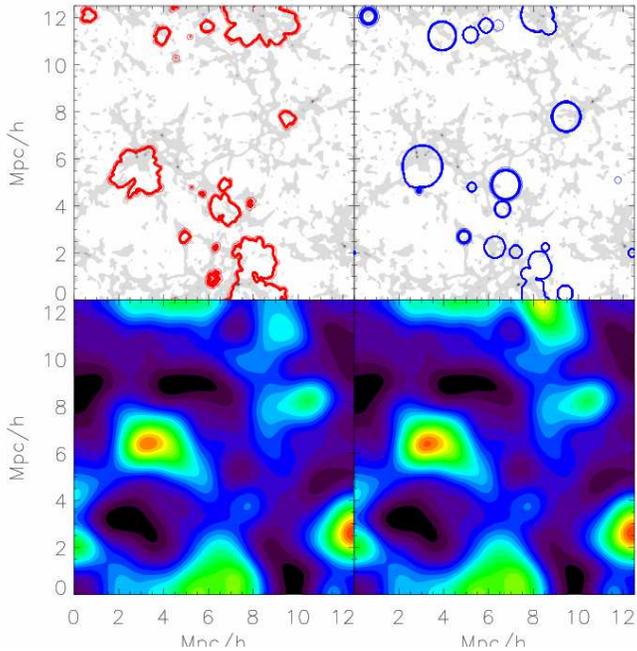}
\vspace{.2cm}
\caption{\emph{The effect of convolution}: The ionization fronts of
  CRASH (red contours, top left) and BEARS (blue contours, top right)
  are overplotted on the underlying density field shown in light
  grey. This slice is extracted from a $256^3$ box at redshift
  six. The corresponding figures below show the images after being
  smoothed by the beam response of the antenna. Details of the
  instrument are given in \S\ref{sec:instrumental}. As expected, the
  images look almost identical after the convolution operation.}
\label{fig:compconv}
\end{figure}

Although there are still differences in the images at lower redshifts
due to the overlap, we expect that convolving the image with the beam
response of the antenna will result in very similar images. This is
indeed what we see in Fig.~\ref{fig:compconv}.  The slice used in the
figure was obtained from a $256^3$-box at redshift 6.  The beam
response (see \S\ref{sec:instrumental}) eliminates the detailed
structures of the ionizing front tracked by the \hbox{3-D} radiative
transfer code.

\subsection{Statistical comparison}

In this section we describe several statistical comparisons between
the results obtained by CRASH and BEARS. The bottom panel of
Fig.~\ref{fig:overlap9withmem} shows the difference between the
mass-weighted mean and variance of the neutral fraction in the
$12.5\cMpch$, $128^3$ box as a function of redshift including the reionization history. The difference is within 2\% for high redshifts
($z>9$), and even at redshift six it is around 5\%.

\begin{figure}
\centering
\hspace{-1cm}
\includegraphics[width=.5\textwidth]{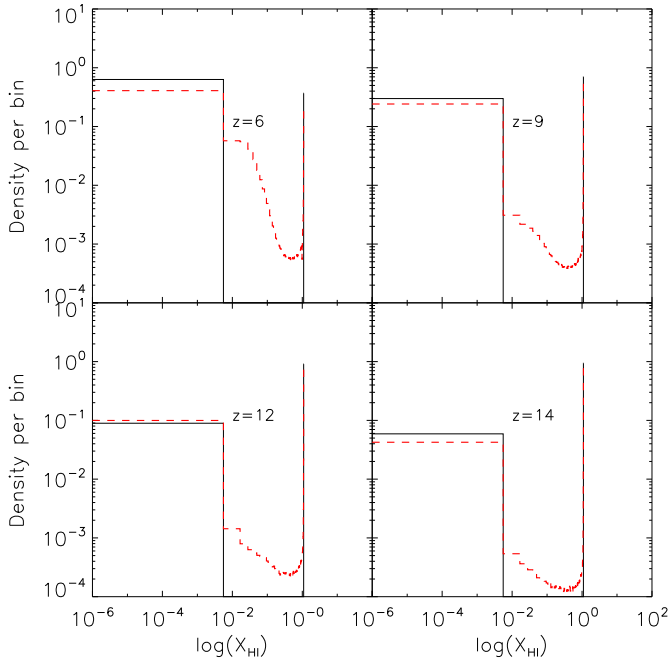}
\vspace{.2cm}
\caption{Histogram of the fractional volume at various average neutral
fractions of hydrogen in a $12.5\cMpch$, $128^3$ box at four different
redshifts as indicated. The black solid line corresponds to BEARS and
the dashed red line to CRASH.}
\label{fig:comphist}
\end{figure}

As a second probe, Fig.~\ref{fig:comphist} shows a histogram of the
fractional volume in a $12.5\cMpch$, $128^3$ box occupied by different
neutral fractions. This plot reveals details of the discrepancy in the
two approaches. At higher redshifts the volume is predominantly
neutral, whereas at lower redshifts, as reionization proceeds,
radiation (partially) ionizes parts of the volume. We see that the
black solid line in the figure, which corresponds to BEARS agrees very
well with the histogram of CRASH (red-dashed) at very low
($X_{HI}<10^{-3.5}$) and at very high ionization levels ($X_{HI}
\approx 1$), but during the intermediate ionization levels they do not
compare very well. The reason for this is that the ionized bubbles in
BEARS have sharp transitions from neutral to a fixed ionized fraction
at the ionization front whereas in CRASH, depending on the density
distribution, the ionized fraction across the ionization front falls
off slowly.

\begin{figure}
\centering
\hspace{0cm}
\includegraphics[width=.5\textwidth]{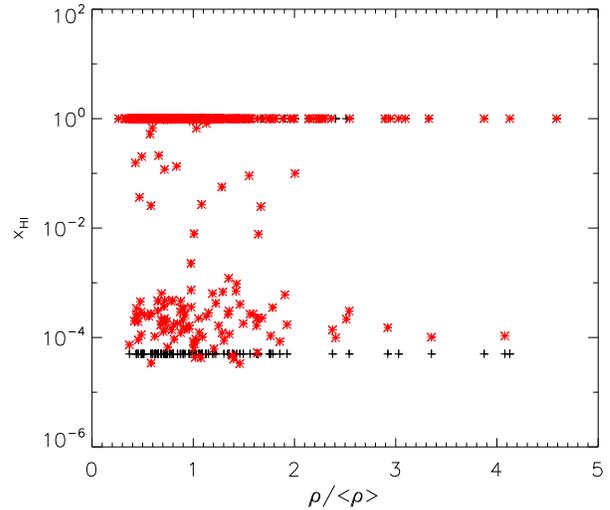}
\vspace{.2cm}
\caption{The above diagram depicts the neutral fraction at a given
  voxel in the simulation box as a function of the
  overdensity at that pixel (only a randomly chosen subsample of all
  the voxels are shown). In the above figure points corresponding to
  CRASH (red) span a larger range of ionizations compared to BEARS
  (black) which are more clustered. Results plotted are for a
  $12.5\cMpch$, $128^3$ box at redshift 6, including the entire 
  reionization history.}
\label{fig:densVionfrac}
\end{figure}

A final diagnostic is presented in Fig.~\ref{fig:densVionfrac}. The
neutral fraction in each voxel \footnote{ \emph{voxels} is a
short-hand to denote a 3-dimensional pixel. It is a portmanteau of
words, \emph{volumetric } and \emph{pixels}.} is plotted against the
overdensity in that pixel. As in the previous diagnostic there is
overlap between the two methods at high and very low neutral
fractions. But we see that CRASH (red) spans the entire range of
neutral fractions whereas the points corresponding to BEARS (black)
are clustered. This again can be attributed to the sudden transition
in the neutral fraction around radiative sources used by BEARS. Also,
most higher density environments are ionized by BEARS because in
most cases a source is centred on the high density pixels of the box.
Another aspect of the implementation of the BEARS scheme is apparent
in this plot, i.e., at higher densities ($\frac{\rho}{<\rho>} > 2.0$)
the agreement between CRASH and BEARS is extremely good. This is
because sources are usually located at overdense location and BEARS
uses the density around the source location to estimate the radius of
ionized sphere.  Thus, around the source and consequently in high
dense regions, BEARS correctly estimates the radius of the ionized
spheres but fails to do so in the low density enviroment.

\section{Creating the ``frequency cube''}
\label{sec:datacube}

Outputs from different frequency channels of an interferometre/LOFAR
measure the radiation (in our case the 21cm emission) from varying
redshifts. Thus, the spectral resolution of the telescope dictates the
scales over which structures in the Universe are smoothed or averaged
along the redshift direction. The shape of the power spectrum and the
characteristics of the line of sight signal are influenced by this
operation. In order to study its effect on the ``true'' underlying
signal it is therefore imperative to create, from outputs of different
individual redshifts, a contiguous data set of maps on the sky at
different frequencies. This section explains how we create just such a
frequency cube (FC). For the purpose of the LOFAR-EoR experiment we
create a cube spanning the observing window of the experiment, i.e.,
from redshift 6 ($\approx$ 200 $\mathrm{MHz}$) to 11.5 ($\approx$ 115 $\mathrm{MHz}$).

The following steps were employed to obtain maps of the 21cm EoR
signal at different redshifts, given a number of snapshots from a
cosmological simulation between the redshifts of interest:

\begin{itemize}
\item Inputs: Start and End frequencies, $\nu_\rmn{start}$ \&
$\nu_\rmn{end}$ (corresponding to redshifts bounded by 6 \& 12); the
resolution in frequency $\delta \nu$, at which the maps are
created. We typically choose a value $\delta \nu$ smaller than the
bandwidth of the experiment, because once this data set has been
created, we can always smooth along frequency to any desired
bandwidth.

\item Create an array of frequencies and redshifts at which the
maps are to be created,
  \begin{equation}
  \nu_i = \nu_\rmn{start} + i \delta \nu,
  \end{equation}
where $i=0, 1, \ldots, (\nu_\rmn{end}-\nu_\rmn{start})/\delta \nu$, 
  and convert each of these frequencies into a redshift array,
  \begin{equation}
  z_i = \frac{\nu_\rmn{21}}{\nu_i} - 1,
  \end{equation}
 where $\nu_\rmn{21}= 1420 ~\rmn{[\mathrm{MHz}]}$, the rest frequency of the 21cm
 hyperfine hydrogen line.
 
\item Now, for each of these redshifts $z_i$, we identify a pair of
simulation snapshots whose redshifts bracket $z_i$. Let us call these
snapshots $z_L$ and $z_H$, such that $z_L < z_i < z_H$.

\item Although all voxels in the output of a simulation
  are at the same redshift ($z_L$ in this case), we assume that the
  voxels of the central slice (along the redshift or frequency
  direction) correspond to the redshift of the simulation box. We then
  move to a slice, $S_\rmn{L}$ (we wrap around the box if necessary,
  since the simulations are performed with periodic boundary
  conditions) at a distance corresponding to the radial comoving
  distance between $z_L$ and $z_i$ given by;
\begin{equation}
D_\rmn{z_L\rightarrow z_i} = \frac{c}{H_o}\int_\rmn{z_L}^{z_i} \frac{dz}{(1+z)\sqrt{X(z)}},
\end{equation}
with $X$ defined as;
\begin{equation}
X(z)=\Omega_m(1+z) + \Omega_r(1+z)^2 + \Omega_l/(1+z)^2 + (1 - \Omega_\rmn{tot}).  
\end{equation}
We use parameters obtained from \emph{WMAP3} (\citealp{Spergel:2007}).

\item With slice $S_\rmn{L}$ as the centre, we average all the slices
within $\pm \Delta z/2$, where $\Delta z$ is the redshift interval
corresponding to a frequency width of $\Delta \nu$ at redshift
$z_L$. Lets call this average $SL_\rmn{avg}$.

\item All the steps performed above do not account for the
time-evolution of the slice between the two outputs and thus it is
important to interpolate between two corresponding slices (to preserve
the phase information of the underlying density field). Therefore, we
identify the corresponding slice, $S_\rmn{H}$ (in position w.r.t the
central slice) in the snapshot $z_H$. Again, we average slices within
$\pm \Delta z/2$ with slice $S_\rmn{H}$ as the centre and call it
$SH_\rmn{avg}$. Note, however, that the number of slices in the
previous step need not be the same as in this step, owing to the fact
that the $\Delta z$ corresponding to $\Delta \nu$ is a function of
$z$.

\item The final step in the generation of the slice $S_{z_i}$, at $z_i$,
involves the interpolation of $SL_\rmn{avg}$ and $SH_\rmn{avg}$:
\begin{equation}
S(z_i) = \frac{(z_i-z_L)SH_\rmn{avg}+(z_H-z_i)SL_\rmn{avg}}{z_H-z_L}.
\end{equation}

\end{itemize}

\begin{figure*}
\includegraphics[width=1.0\textwidth]{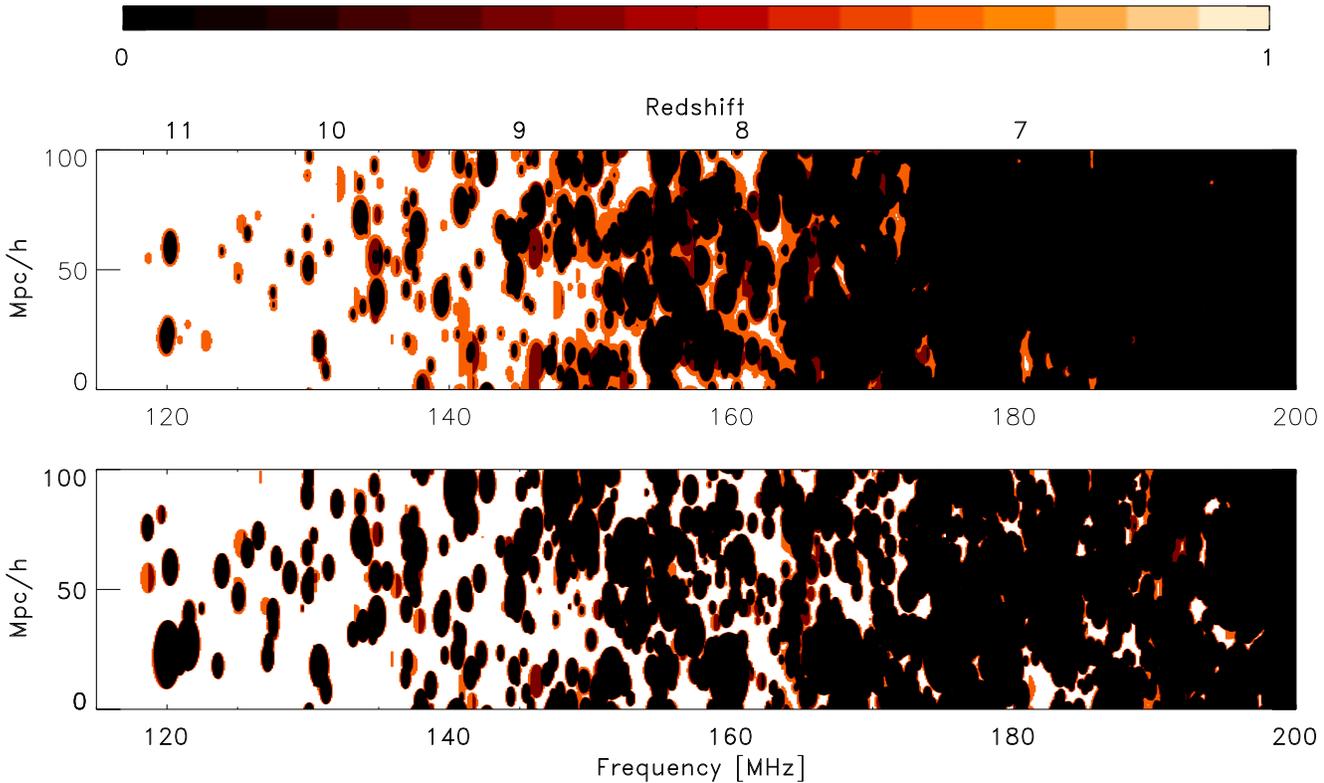}
\caption{ A slice along the frequency direction for the neutral
fraction for two different scenarios. One with quasars (top panel) and
the other with stars (bottom panel). The repetition along the
frequency direction, which is expected due to the finite size of the
box, is reduced by a factor $\sqrt{3}$ because the slice is obtained
diagonally across the box.  We see from the figure that although the
stars do start ionizing earlier, for the models we have developed
(details in the text, section \ref{sec:starsVquasars}), quasars are
far more efficient at ionizing the IGM.  }
\label{fig:alongfreqIon}
\end{figure*}

Fig.~\ref{fig:alongfreqIon} shows an example of a slice through the
box for the the ionization history due to stars constructed using the
algorithm above. Because the size of the box used is
$100~h^{-1}~\rmn{Mpc}$ in comoving coordinates and the comoving radial
distance between redshifts of 6 and 12 is
$\approx1600~h^{-1}~\rmn{Mpc}$, we would expect there to be repetition
of structures along the frequency/redshift direction. In plotting the
above figure this has been reduced by a factor $\sqrt{3}$ since the
slice has been extracted along the diagonal of the FC. In the
following sections we will do all our analysis and comparisons on the
box generated in the above manner.

Although the radiative transfer is performed on each box with the
 underlying density in real space, we produce the FC for the
brightness temperature with densities estimated in redshift
space to account for the redshift distortions introduced by the peculiar
velocities.

\section{Stars versus Quasars: exploring different source-scenarios}

\label{sec:starsVquasars}
  
 Although the general consensus is that the dominant sources of
 reionization are stellar in nature, there is also evidence to support
 quasars as additional sources of reionization
 \citep{kuhlen05,rajat08,zaroubi07}. One of the main differences
 between stars and quasars, apart from the extent of reionization
 caused by individual sources, is the heating due to quasars. We
 simulate only the reionization history for the cases involving either
 stars or quasars and leave simulating the effect of heating for
 future work. We use a $512^3$ particle, $100~h^{-1}$ Mpc ($L_{box}$)
 dark-matter-only simulation for this purpose. For details on the
 N-body simulation see Section \ref{sec:nbody}. About 75 simulation
 snapshots between redshifts 12 and 6 were used to create the FC. The
 entire operation of doing the radiative transfer on each of these
 snapshots and converting them into the FC takes approximately 12
 hours on an 8-dual-core AMD processor machine with 32GB of memory.

 In the sections below we describe the prescriptions used to embed the
 dark matter haloes with quasars and stars. Following that we discuss
 some of the statistical differences in the ionization fractions and
 contrast the statistical nature of the $\delta T_b$ for these
 different scenarios, as a function of frequency. We caution that the
 scenarios described here are mostly meant to illustrate the potential
 of the techniques being developed here. For example, although the
 quasar model described below seems to ionize the Universe
 effectively, care has not been taken to constrain the quasar
 population based on the soft X-ray background excess between 0.5 and
 2 KeV \citep{dijkstra}.

\subsection{Prescription for \emph{quasar type} sources}

 From the output of the N-body simulation, dark-matter haloes were
identified using the friends-of-friends algorithm and a linking length
$b=0.2$. Black holes embedded in them were assigned a mass
according to \citet{zaroubi07},

\begin{equation}
 \mathrm{M_{BH}}= 10^{-4}\times\frac{\Omega_b}{\Omega_m}M_{halo},
\label{eq:mbh}
\end{equation}
 where the factor $10^{-4}$ reflects the Magorrian relation between
the halo mass ($\rmn{M_{halo}}$) and black hole mass ($\rmn{M_{BH}}$),
and $\frac{\Omega_b}{\Omega_m}$ gives the baryon ratio \citep{Ferrarese02} .

 The template spectrum we assumed for the quasars is a power law of the form,
\begin{equation}
F(E) = {\cal A}\;E^{-\alpha} \,\,\, \,\,\, 10.4 \mathrm{eV} < E < 10~
\mathrm{keV},
\end{equation}
where $\alpha$ is the power-law index which is set to unity.
A quasar of mass $M$ shines at $\epsilon_{rad}$ times the Eddington
luminosity,
\begin{equation}
L_{Edd}(\mathrm{M_{BH}}) = 1.38\times10^{38} \left(\frac{\mathrm{M_{BH}}}{\mathrm{M_\odot}}\right) [\rm erg \; s^{-1}].
\end{equation}
Therefore ${\cal A}$ is given by:
\begin{equation}
 {\cal A}(\mathrm{M_{BH}}) =\frac{ \epsilon_{rad}\; L_{Edd}(\mathrm{M_{BH}})}{\int \limits_{E_{range}}
E^{-\alpha}\;dE \times 4\pi r^2 } \;[\rm erg s^{-1}cm^{-2}], 
\end{equation}
where  $E_{range} = 10.4 \mathrm{eV}\, - \, 10 ~\mathrm{keV} $. 
 
 In the above model, we have assumed that the spectral energy distribution
(SED) of the quasar extends well above the ionization energy of
hydrogen. This ensures that copious amounts of photons are available
to ionize the 
hydrogen in the IGM while at the same time reducing the number of hard
X-ray photons that could potentially heat the IGM. Because we 
concentrate on the ionization history in this paper, we postpone a
detailed study of the effect of cutoff in the SED at the higher and lower
end of the energies.

\subsection{Prescription for \emph{stellar} sources}
 
We associate stellar spectra with dark matter halos using the
following procedure. The global star 
formation rate was calculated using;

\begin{equation}
\dot{\rho_\star}(z) = \dot{\rho_m} \frac{\beta\exp\left[\alpha(z - z_m)\right]}{\beta - \alpha + \alpha\exp\left[\beta(z - z_m)\right]}~~[{\rm M}_\odot~{\rm yr}^{-1}~{\rm Mpc}^{-3}],
\label{eq:starformrate} 
\end{equation}
where $\alpha= 3/5$, $\beta=14/15$, $z_m=5.4$ marks a break redshift,
and $\dot\rho_m= 0.15\,{\rm M}_\odot~{\rm yr}^{-1}{\rm Mpc}^{-3}$
fixes the overall normalisation \citep{springel03} .  Now, if $\delta
t$ is the time interval between two outputs in years, the total mass
density of stars formed is 
\begin{equation}
\rho_\star (z)\approx \dot\rho_\star(z)\delta t ~~[{\rm M}_\odot~{\rm Mpc}^{-3}].
\end{equation} 
Notice that this approximation is valid only if the typical lifetime
of the star is much smaller than $\delta t$, which is the case in our
model because we assume 100$M\odot$ stars as the source that have a
lifetime of about few Myrs \citep{schaerer}.

Therefore, the total mass in stars in the box is $ M_\star(box)
\approx L_{box}^3 \rho_\star ~~[{\rm M}_\odot]$. This mass in stars is
then distributed among the halos according to the mass of the halo as,
\begin{equation}
m_\star(halo) = \frac{m_\rmn{halo}}{M_\rmn{halo}(tot)} M_\star(box),
\end{equation}
where $m_\star(halo)$ is the mass of stars in ``halo'', $m_\rmn{halo}$
the mass of the halo and $M_\rmn{halo}(tot)$ the total mass of halos
in the box.

We then assume that all of the mass in stars is distributed in stars
of 100 solar masses, which implies that the number of stars in the
halo is $N_{100} = 10^{-2} \times m_\star(halo)$. The number of
ionizing photons from a 100 $M_\odot$ star is taken from Table 3 of
\citet{schaerer} and multiplied by $N_{100}$ to get the total number
of ionizing photons emanating from the ``halo'' and the radiative
transfer is done assuming these photons are at $13.6eV$. The escape
fractions of ionizing photons from early galaxies is assumed to be
10\%.

\subsection{Statistical differences in the history of reionization}

Using the models of ionizing sources and the algorithm to generate
the FC described above, we plot a slice along the frequency
direction of the ionization histories in Fig.~\ref{fig:alongfreqIon}.
We see from the figure that the stars start ionizing much earlier than
the quasars, but the rate at which they ionize is low. Therefore,
even though the effect of quasars on the ionization is seen later, they
manage to ionize the IGM earlier than stars. This is mainly due to the
efficiency (per unit baryon) of the quasar to ionize the medium and
to the fact that in the model we have constructed the miniqso population
is not constrained based on the excess of soft X-ray background
radiation and therefore the miniqso number density is higher than it
would be in reality. It should be noted that if the escape fractions
are allowed to change the ionization histories due to stars can be
altered significantly.

\begin{figure}
\includegraphics[width=0.5\textwidth]{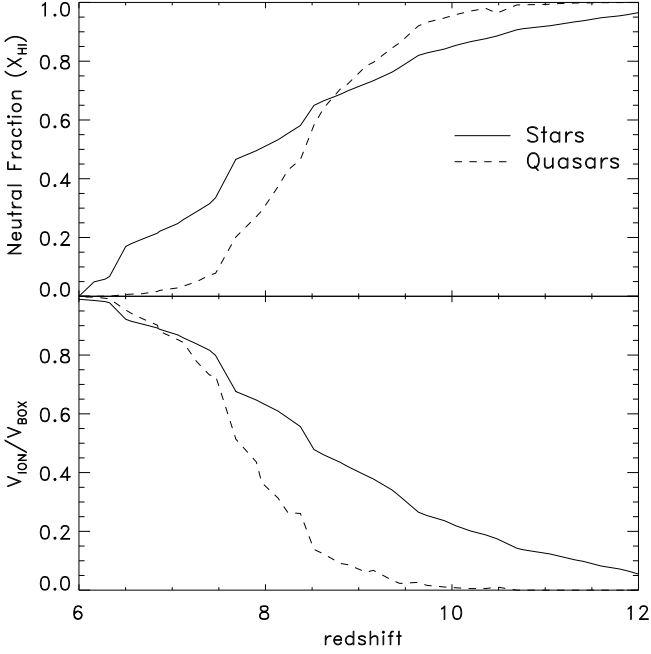}
\caption{The figure shows the volume averaged mean neutral fraction
(top panel) in the box as a function of redshift for stars (solid) and
quasar (dashed) populations. Although both the populations ionize the
entire box by redshift 6, the ionization history follows very
different paths. Within the models used for the simulation the quasars
seem to ionize earlier than the stars. A similar diagnostic is the
volume fraction of the box that is ionized as a function of redshift.}
\label{fig:MeanBoxes}
\end{figure}

This can be seen more clearly and quantitatively in
Fig.~\ref{fig:MeanBoxes}, which shows the volume averaged mean neutral
fraction as a function of redshift (top panel) and the volume fraction
that is ionized ($X_{HI}<0.05$) in the box (bottom panel).

\begin{figure*}
\includegraphics[width=1.0\textwidth]{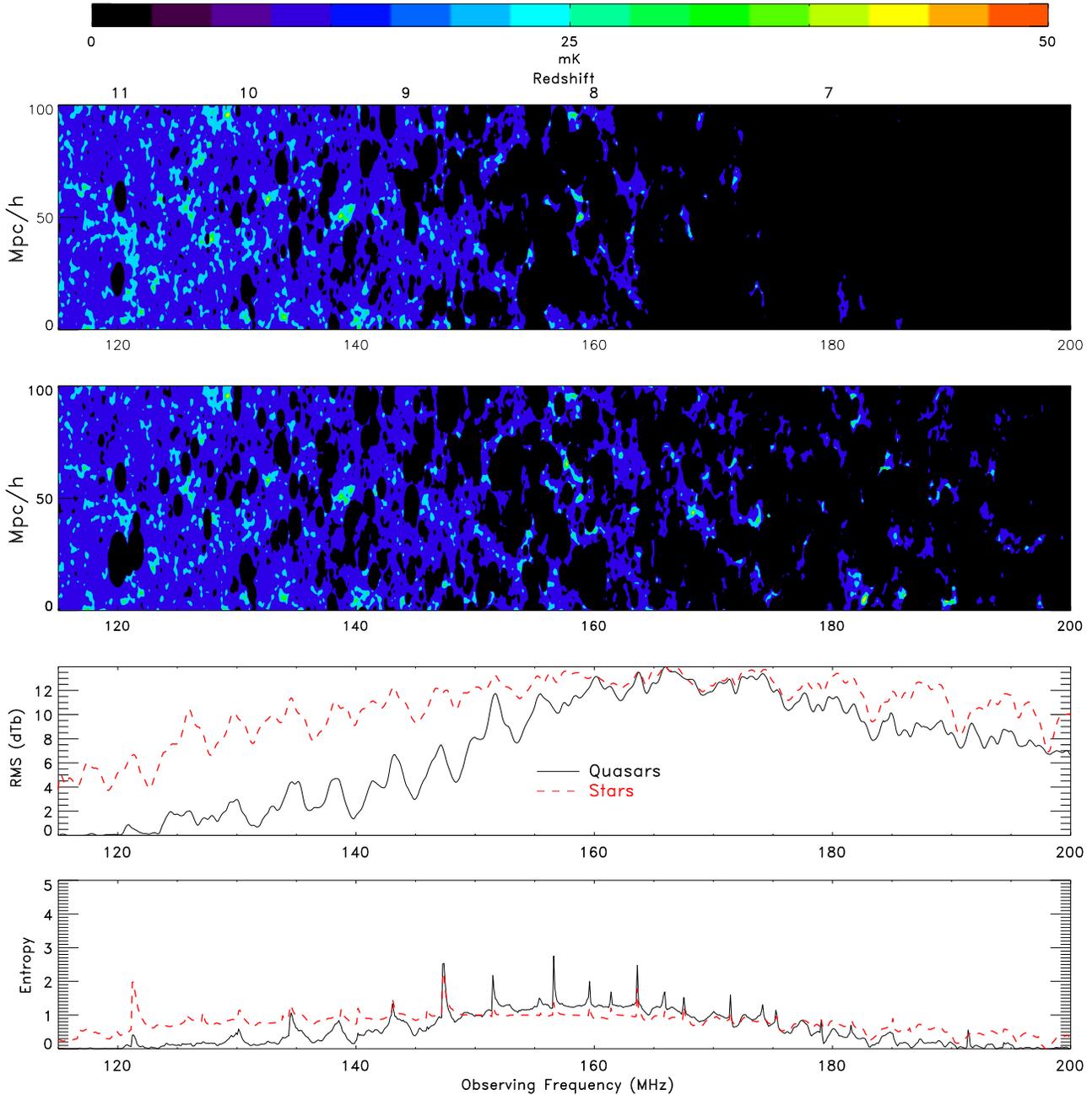}
\caption{Starting from the top, \emph{panel 1} shows the $\delta T_b$
distribution of quasars, \emph{panel 2} shows the same for stars, \emph{panels 3
 and 4} show the variance and entropy of slices at different frequencies for
stars (red dashed) and quasars (black solid). The figure clearly shows that the
 nature of the reionization history differs significantly between
 stars and black holes. In 
the case of stars reionization seems to be much more extended than for the
case of stars.}
\label{fig:alongfreqdtb}
\end{figure*}

Fig.~\ref{fig:alongfreqdtb} shows a slice through the
FC of $\delta T_b$s for the case of quasars (first  
panel from the top) and stars (second panel) which was calculated
following \citet{madau97}: 

\begin{eqnarray}
\delta T_b & =
\left(20~\mathrm{mK}\right)&\left(\frac{X_{HI}}{h}\right)\left(1-\frac{T_{CMB}}{T_{spin}}\right)
\nonumber \\ & &\times \left(\frac{\Omega_b h^2}{0.0223}\right)
\left[\left(\frac{1 +
z}{10}\right)\left(\frac{0.24}{\Omega_m}\right)\right]^{1/2},
\label{eq:dtb}
\end{eqnarray}
where $h$ is the Hubble constant in units of
$100\mathrm{km~s^{-1}~Mpc^{-1}}$, $\delta$ is the mass density
contrast, and $\Omega_m$ and $\Omega_b$ are the mass and baryon
densities in units of the critical density. $T_{CMB}(z)$ is the
temperature of the cosmic microwave background at redshift $z$ and
$T_\rmn{spin}$ the spin temperature. In all our calculations we have
assumed $T_{spin} \gg T_{CMB}$.

The third panel in Fig.\ref{fig:alongfreqdtb} shows the variance of
$\delta T_b$ as a function of frequency for stars (red dashed
line) and quasars (black solid lines). Since quasars start ionizing
late and completely ionize the Universe by redshift 6, the variance
stays below that for stars at the beginning and end of reionization
but peaks at around 160 $\mathrm{MHz}$ corresponding to a redshift of
7.8. Interestingly, for stars the peak of the variance occurs
around the same frequency. This is a welcome coincidence from an
observation point of view. LOFAR will initially observe using
instantaneous bandwidths of 32 $\mathrm{MHz}$ (Lambropoulous et al., \emph{in
  prep}). Thus, it becomes important to make an educated guess of the
frequency around which the observation has to be made, to be sure that
we capture reionization activity at its peak. And because the variance
of these two very different scenarios peak around the same redshift,
this frequency ($\approx 160 \rmn{[MHz]}$) is a reasonable guess. It
is worth emphasizing that these are predictions based on the
simple-minded models prescribed above. Nevertheless this exercise
illustrates the capability of our algorithm to implement diverse
scenarios and help guide the observational strategies of
LOFAR. 

Finally, in the fourth panel we plot the image (slice) entropy
as a function of redshift.  Image entropy is a measure of randomness
in the image and in our case it reflects the inhomogeneity of
reionization as a function of redshift. The entropy of an image is
calculated as follows: first, a histogram of the intensities in the
image, which in our case correspond to $\delta T_b$ values, is made
and normalized. If there are $\rmn{N_{bins}}$ bins and each bin $i$
has a normalized value $\rho_i$, then the entropy is given by,
\begin{equation}
\mathrm{Entropy} = \sum_{\mathrm{N_{bins}}} \rho_i log\rho_i.
\label{eq:entropy}
\end{equation}

 The meaning of image entropy is illustrated with an example in
section \S\ref{sec:instrumental}, aided with
Fig.~\ref{fig:mockentrop}. In the context of our study the image
entropy has the following meaning:
\begin{enumerate}
\item \emph{Before convolution with the LOFAR beam
      (ref. Fig.~\ref{fig:alongfreqdtb})}:\\ At low frequencies ($< 145
      \mathrm{MHz}$) the universe is still mostly neutral and
      homogeneous. Thus a histogram of $\delta T_b$ at a given
      frequency less than 145 $\mathrm{MHz}$ would be very narrow and hence
      have lower entropy (ref. Eq \ref{eq:entropy}).  But as the sources start
      ionizing the IGM, it introduces fluctuations in $\delta T_b$ and
      the histogram spreads out thus increasing the
      entropy. Similarly, at much higher frequencies ($>180 \mathrm{MHz}$) the
      entropy drops down because the Universe is largely ionized and
      the histogram of $\delta T_b$ is narrow but now centred at 0.
\item \emph{After convolution with the LOFAR beam
      (ref. Fig.~\ref{fig:postconvStats})}:\\ In this case, although
      the overall behaviour is similar to that of the previous case
      before convolution, the effect is enhanced. This can be
      understood with the aide of Fig.~\ref{fig:mockentrop}. Lower
      frequencies ($120 < \nu < 145 \mathrm{MHz}$) is analogoues to case (a) \&
      (c) and its corresponding histogram in (Ha) \& (Hc). Because, at
      these frequencies the Universe is still largely neutral with
      just a few sources and the effect of smearing $\delta T_b$ with
      the observing beam spreads the histrogram thus increasing
      the entropy. At intermediate frequencies ($120 < \nu < 180
      \mathrm{MHz}$), the effect of decreasing entropy is analogoues to case
      (b) \& (d) with its corresponding histogram in (Hb) \&
      (Hd). Although the entropy increased from (b) to (d) and also in
      case of $\delta T_b$ after the convolution, the increase is
      relatively less. This is because the smearing action was on
      scales larger than the fluctuations in $\delta T_b$ hence
      narrowing the histogram to a few values. At higher frequencies
      of course $\delta T_b$ is zero and hence the entropy drops to 0
      as well.
\end{enumerate}

A final statistic that was calculated was the probability distribution
function (PDF) of the $\delta T_b$ at four different redshifts for stars and
quasars. If the IGM is not ionized and if the neutral hydrogen is
measurable (i.e.\ the signal-to-noise ratio is high enough), the
distribution of $\delta T_b$ reflects the underlying density 
field (eq.~\ref{eq:dtb}). But as reionization proceeds, larger
and larger regions become ionized and the PDF of $\delta T_b$ skews because
of the excess of $\delta T_b$ close to zero. This fact can be made use of as a
tool for the statistical detection of the EoR signal as will be
explored in an upcoming paper by Harker et al., \emph{in prep}. The features of
the PDF seem to vary dramatically between the two scenarios.

\begin{figure*}
\hspace{-0.7cm}
\includegraphics[width=1.0\textwidth]{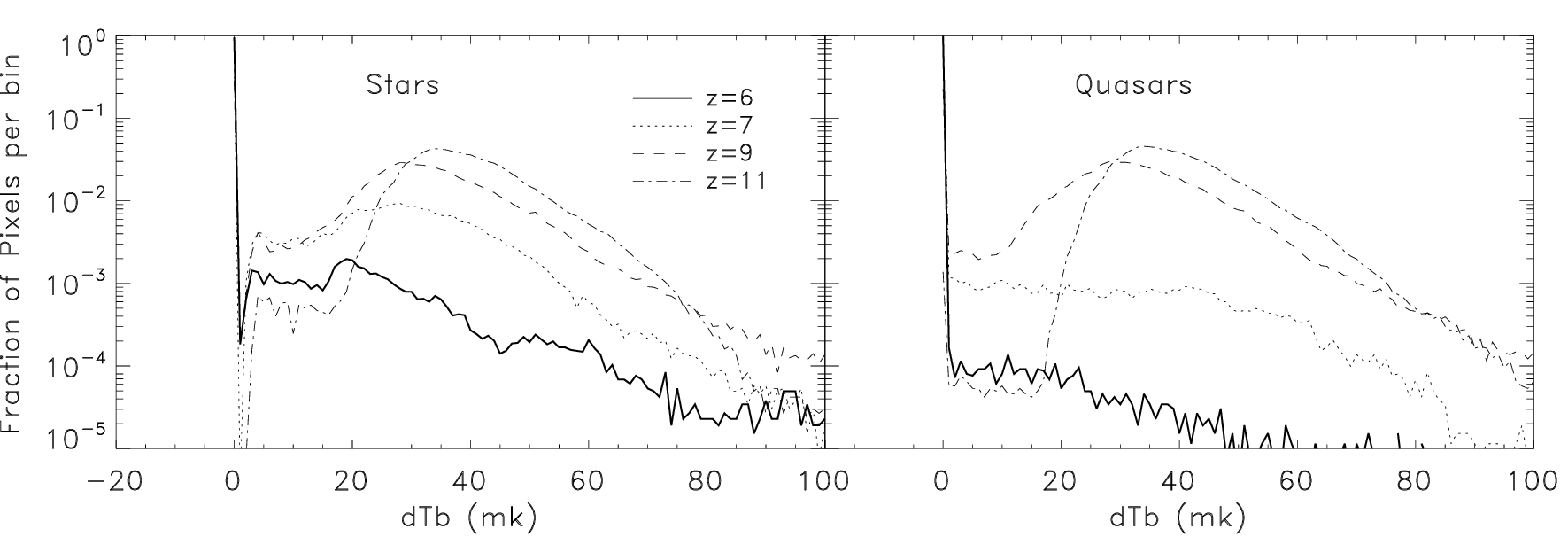}
\caption{The probability distribution functions
(PDF) of $\delta T_b$ at redshifts 6, 7, 9 and 11, as
indicated by the line styles, for stars (left) and quasars (right). The
PDF should reflect the underlying density distribution if the IGM is
neutral, but ionization of the IGM causes a peak at a brightness
temperature of zero skewing the distribution.}
\label{fig:dtb_pdf}
\end{figure*}

\section{LOFAR response and its effects}
\label{sec:instrumental}

The rationale behind producing these simulations is to test the effect
of the interferometric response on the signal and other contributors with
the same frequency band. This section gives an overview of
simulations of the LOFAR response and shows how the signal will be seen
by LOFAR in the absence of noise or other calibration errors. A more
detailed discussion of the LOFAR response and the data model for the LOFAR-EoR
experiment will be provided in Lambropoulous et al., \emph{in
  prep}. For the EoR experiment we plan to use the LOFAR core which
will consist of about 24 dual stations of tiles. Each dual
station will have a diameter of approximately 35 meters and the
maximum baseline between stations will be about 2 km.
 
For the simulations in this paper we make some simplifying assumptions
regarding the telescope response. We assume the narrow bandwidth
condition and we also assume that the image-plane-effects have been
calibrated to a satisfactory level. This includes a stable primary
beam and adequate compensation for the ionospheric effects such as the
ionospheric phase introduced during the propagation of electromagnetic
waves in the ionosphere and the ionospheric Faraday rotation.  In an
interferometric observation, the measured correlation of the electric
fields between two sensors $i$ and $j$ is called visibility and
is given by \citet{perley89}:
 
\begin{equation}\label{vis}
  V(u,v,w)=\int A(l,m,n)I(l,m,n)e^{-2\pi i(ul+vm+wn)}\mathrm{dl} \mathrm{dm} \mathrm{dn},
\end{equation}
 where $A$ is the primary beam, $I$ the sky brightness, $l,m,n$ are the
direction cosines and $u,v,w$ are the coordinates of the baseline (in
units of observing wavelength) as seen from the source.
 
We further assume that simulated maps are a collection of point
sources: each pixel corresponds to a point source with the relevant
intensity. Note that the equation above takes into account the sky
curvature. The visibilities are sampled for all station pairs, but also
at different pair positions as the Earth rotates.
 
For every baseline and frequency the \emph{uv}\footnote{The \emph{uv}
  plane is the Fourier-pair of the sky brightness, where \emph{u} and
  \emph{v} are the distances between antenna pairs on an
  interferometer measured in terms of the wavelength of observation.}
track points sample different scales of the Fourier transform of the
sky at that frequency. We compute the \emph{uv} tracks for each
interferometer pair for 6 hours of synthesis with an averaging
interval of 100 sec and we grid them onto a regular grid in the
\emph{uv} plane.  Using the gridded tracks as a sampling function, we
sample the Fourier transform of our model sky, which in our case is
the 21cm EoR signal, at the corresponding grid cells. This gives us
the time series of the visibilities for every station pair.
 
The Fourier (or Inverse Fourier) transform of the sampled visibilities
is called the "dirty" map. It is actually the sky map convolved with
the Fourier transform of the sampling function, which is called the
"dirty" beam or the PSF. This is a simple-minded approach in estimating
the sky brightness as it uses linear operations. 
 
With the LOFAR core we expect to reach a sensitivity of 350~mK in one
night of observations at 150 $\mathrm{MHz}$. After 100 nights the final
sensitivity will drop to 35mK. For total intensity observations we
expect Gaussian noise with zero average and the above mentioned rms to
drop to 35~mK. We assume that calibration errors, after 100 nights of
integration accumulate in such a way that they follow a
Gaussian distribution obeying the central limit theorem.

\begin{figure*}
\includegraphics[width=1.0\textwidth]{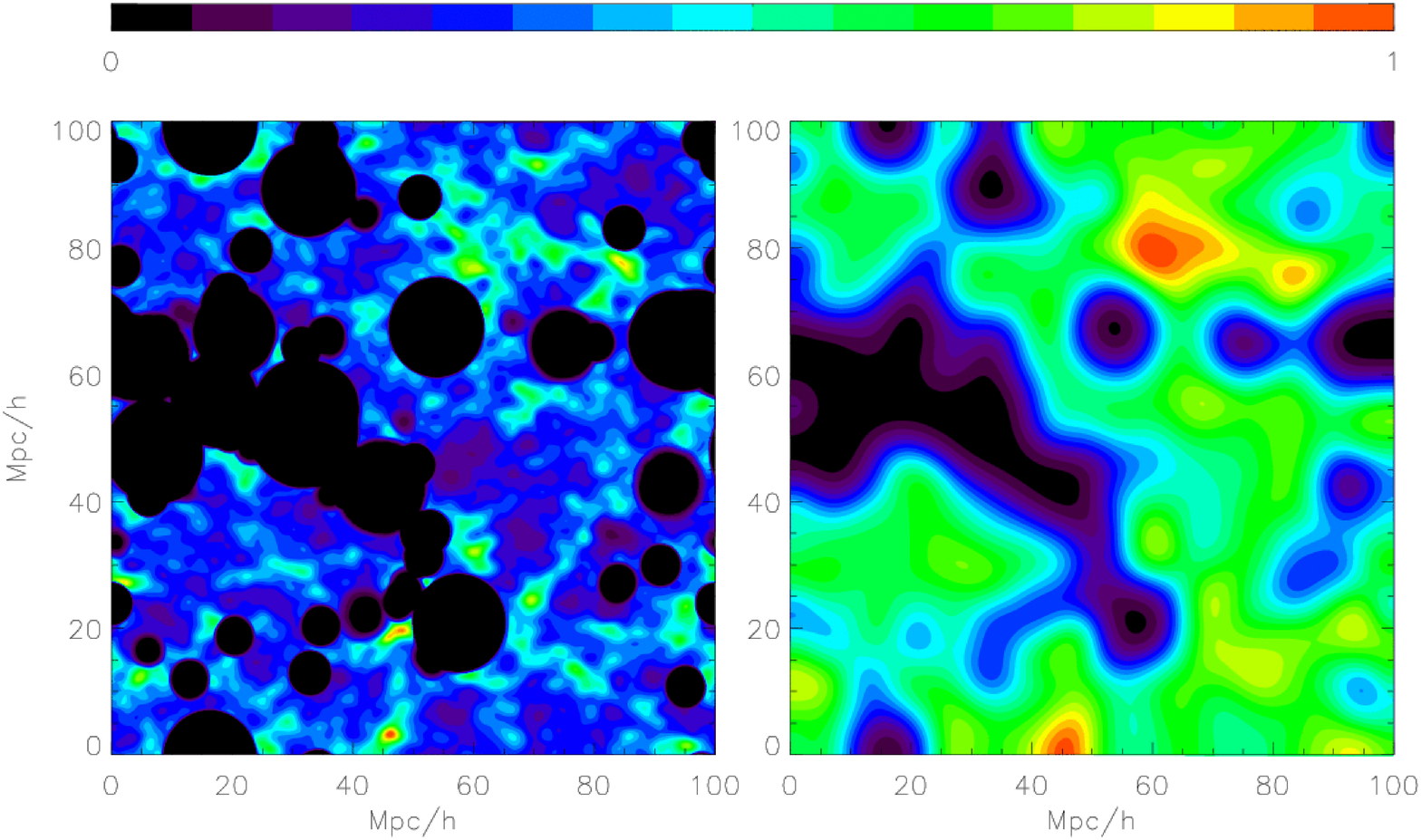}
\caption{A comparison of the brightness temperature (normalized to the
  largest value) between the ``original'' image (left) and the same
  image after convolving with the LOFAR PSF (right) at a frequency
  corresponding to redshift 10.  Since the PSF of LOFAR essentially
  performs as a low-pass filter the convolved image is smoother.  }
\label{fig:dirty_clean}
\end{figure*}

The effect of the ``original'' image passing through the response of
the LOFAR antenna at a frequency of 130 $\mathrm{MHz}$ is shown  in
Fig.~\ref{fig:dirty_clean}. The figure shows the brightness
temperature in both cases normalized to the largest value in the
corresponding figure. Note that the simulation is
$100 \cMpch $ across which is only about ${1/8}^{th}$ of the total
($5^o \times 5^o$) field of view of LOFAR. Therefore we are missing
the very large scale Fourier modes. We are planning to run larger
simulations to overcome this problem. 

\begin{figure*}
\includegraphics[width=1.0\textwidth]{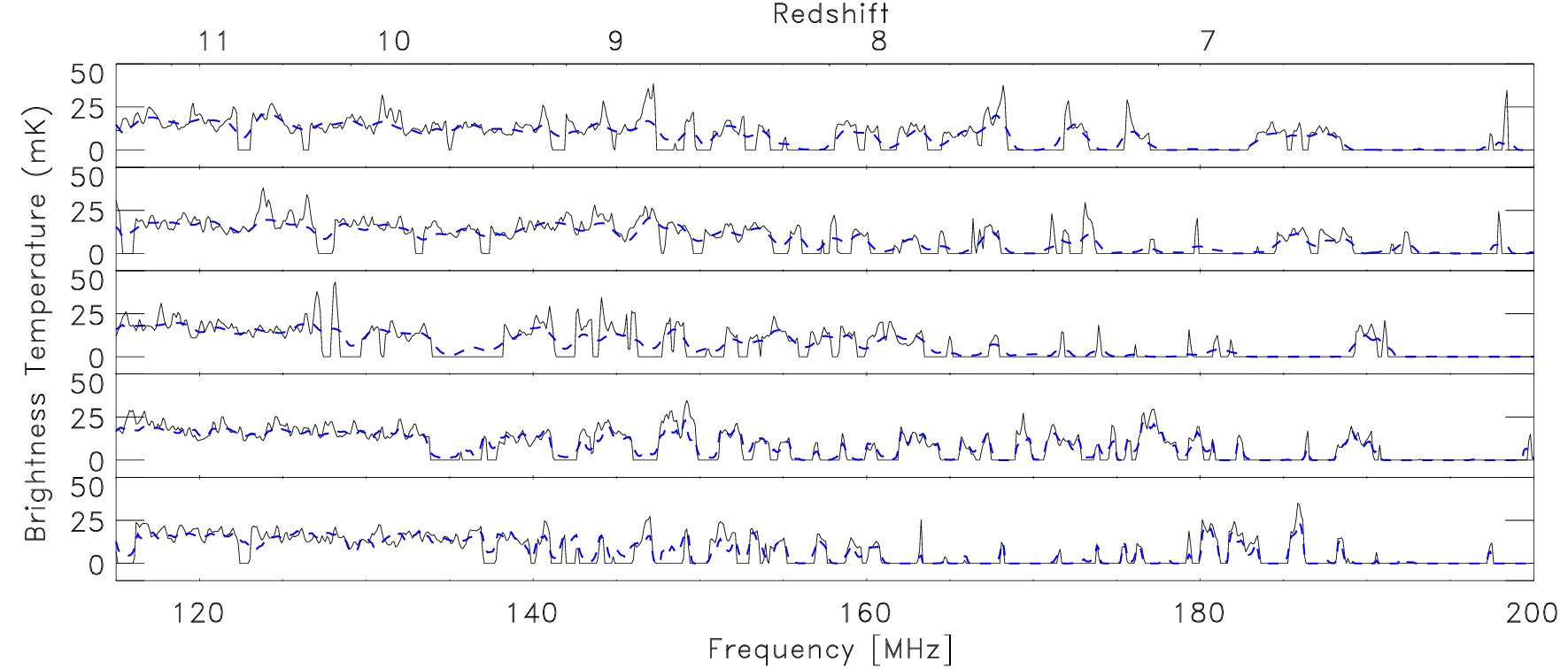}
\caption{Examples of five lines of sight through the frequency
direction for $\delta T_b$ for the case of ionization by
stars. Over-plotted in blue are the lines of sight as observed through 
the LOFAR telescope with a spectral resolution of 1 $\mathrm{MHz}$}
\label{fig:losdtb_stars}
\end{figure*}

\begin{figure*}
\includegraphics[width=1.0\textwidth]{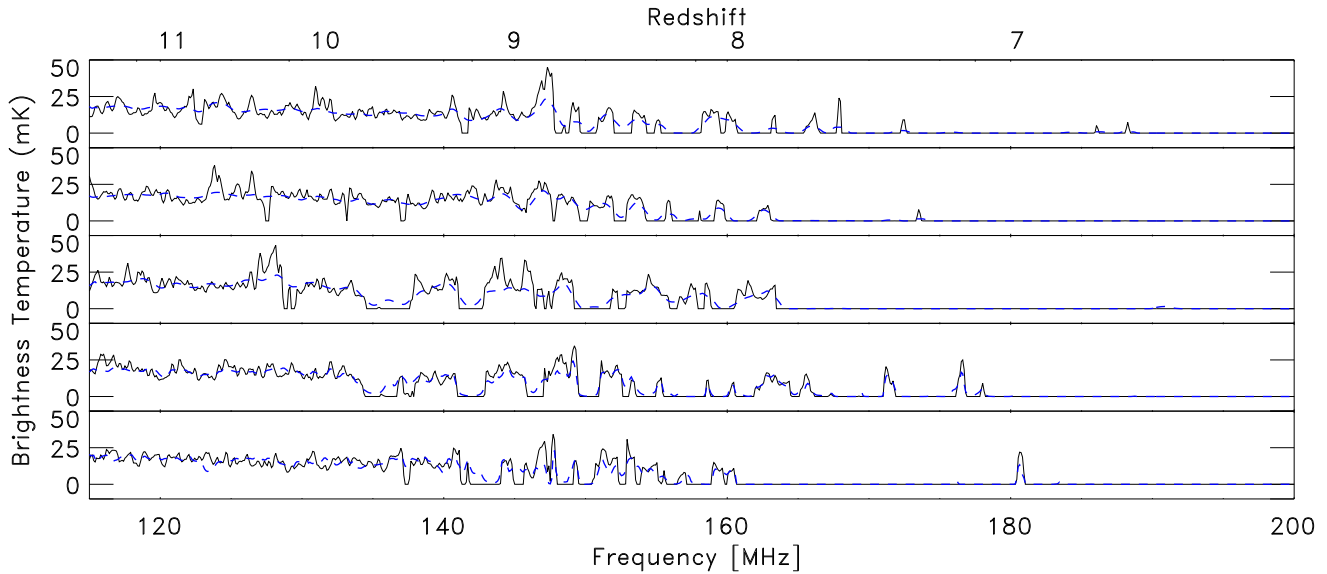}
\caption{Same as Fig.~\ref{fig:losdtb_stars} but for the case of 
ionization by quasars. }
\label{fig:losdtb_qsos}
\end{figure*}

Along with the effect of the spatial resolution of LOFAR on the sky,
we also studied the effect of its spectral resolution.  The maps
produced by LOFAR will have a bandwidth of about 1 $\mathrm{MHz}$.  This is
essential to beat the noise level down. Figs.~\ref{fig:losdtb_stars}
and \ref{fig:losdtb_qsos} show $\delta T_b$ along five different lines
of sight. The black solid lines indicate the signal before convolution
and spectral smoothing, whereas the over-plotted blue dashed lines
include the effect of LOFAR's spatial and spectral response.

\begin{figure*}
\includegraphics[width=1.0\textwidth]{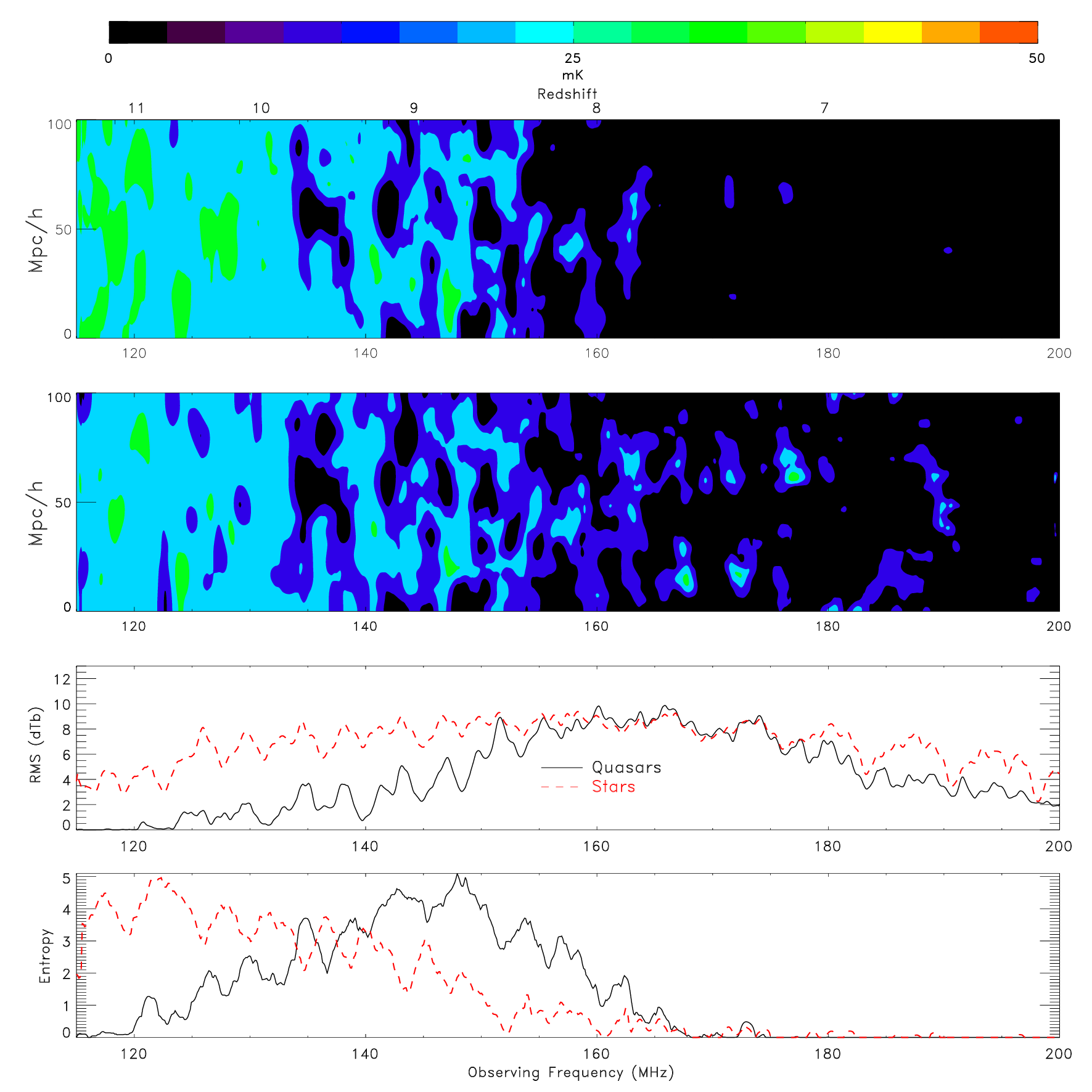}
\caption{Same as in Fig.~\ref{fig:alongfreqdtb} but after the signal has
been passed through the LOFAR response.}
\label{fig:postconvStats}
\end{figure*}

The entire FC was convolved with the antenna response of LOFAR
calculated separately for each frequency. This ``convolved cube'' is
then used to plot Fig.~\ref{fig:postconvStats}. This figure is
identical in structure to Fig.~\ref{fig:alongfreqdtb}. As expected,
due to the smoothing action of the antenna beam pattern, the variance of
the signal in the ``convolved cube'' drops by approximately 30\%
relative to that of the FC while conserving the overall behaviour
across all frequencies. The entropy in each of these slices on the
other hand shows a dramatically different behaviour. The rise and fall
of the entropy in the case of quasars is much more pronounced for the
``convolved cube'' and the entropy for stars shows a steady
increase. It is interesting that the slice-entropies for stars and quasars
are very different. This could be used as a discriminant for different
scenarios of reionization.

\begin{figure}
\hspace{-1.cm}
\includegraphics[width=0.5\textwidth]{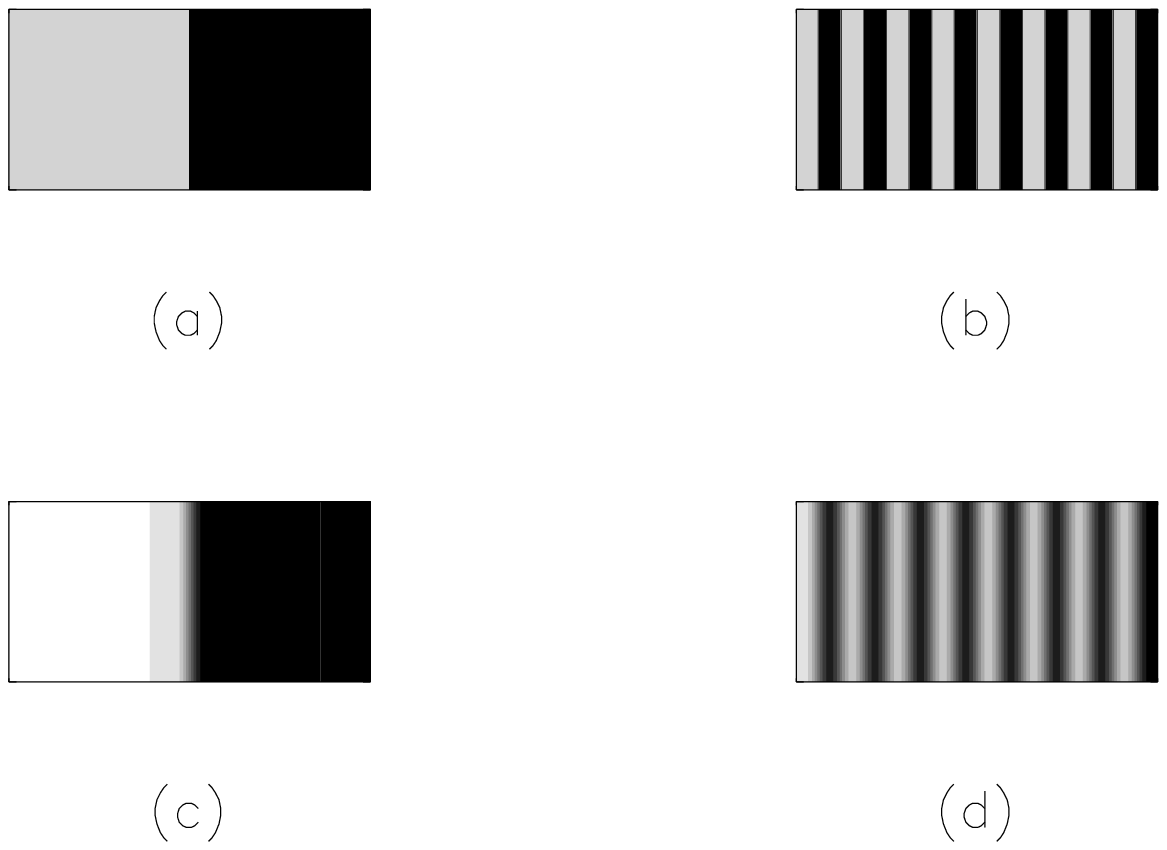}
\includegraphics[width=0.5\textwidth]{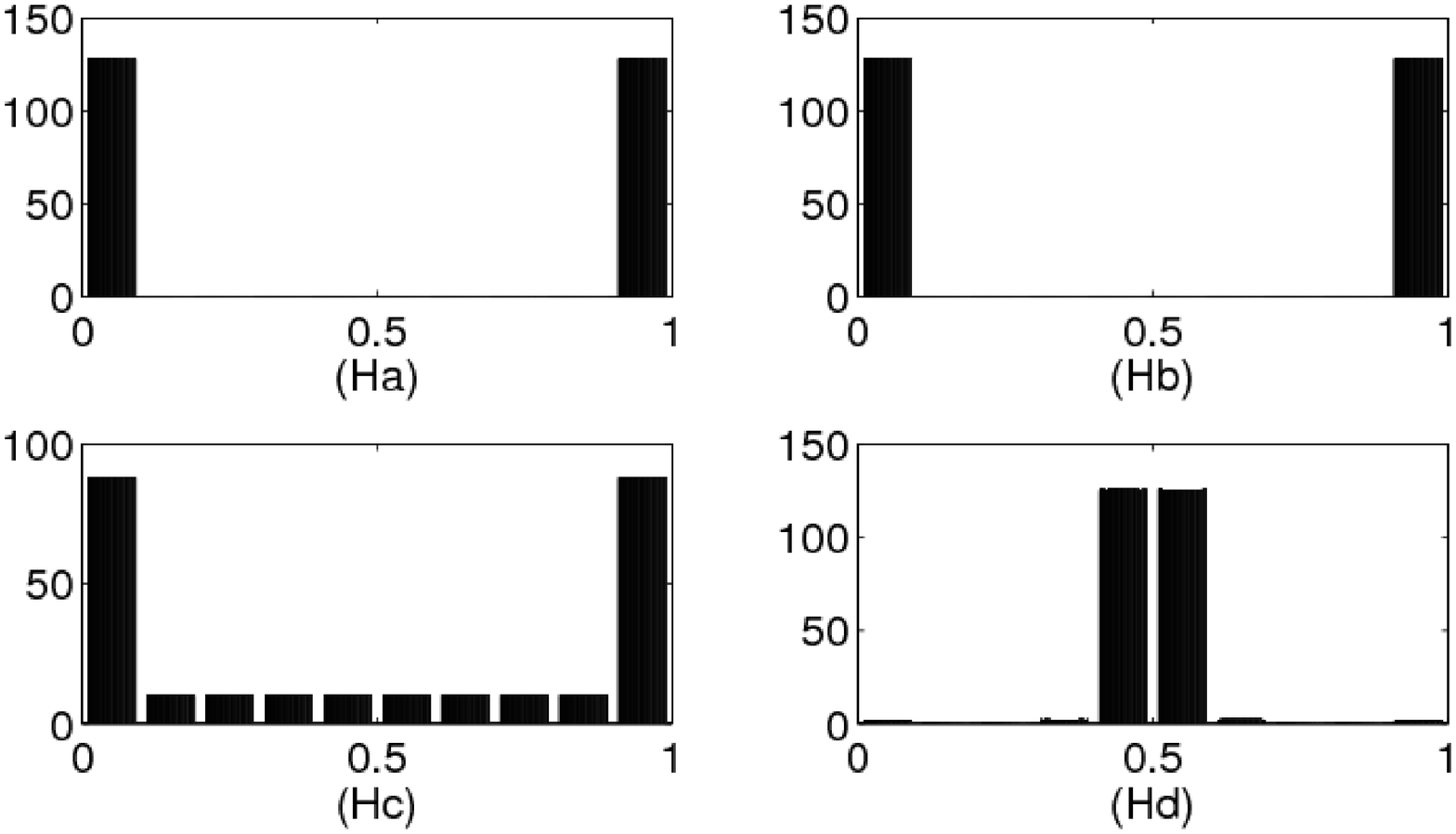}
\caption{Illustration of image entropy: In the above example, images
(a) and (b) have the same image entropy with a value of one according
to equation \ref{eq:entropy}.  Images (c) and (d) are the convolutions
of (a) and (b) with a square top hat function with side of length
$1/4$ of the side of the image. The entropies of these images are 4.1
and 2.2, which is larger than the entropies we started
with. Therefore, when we introduced correlations between pixels by
smoothing them, their entropies increased. Images (Ha) to (Hd) are the
histograms corresponding to images (a) to (d) respectively. As we can
see, the histograms of images (a) and (b) have the same distribution,
whereas image (Hc) has a histogram that has a large spread compared to
(Hd). From an information theory point of view this large spread
corresponds to higher information content and hences higher entropy.
Similarly, when we smooth/average along the frequency direction in the
FC, we introduce correlations between pixels, hence increasing the
entropies considerably when the scale of smoothing is smaller than the
typical correlation length of the pixels in the original image. }
\label{fig:mockentrop}
\end{figure}

Fig.~\ref{fig:mockentrop} is a simple example to illustrate the fact
that smoothing increases the entropy in the figure. In essence, the
averaging along the frequency direction introduces correlations
between pixels of adjacent slices. For the simulations considered
here it corresponds to 10 adjacent slices (since the resolution of the
simulation is 0.1 $\mathrm{MHz}$ and that of LOFAR is 1 $\mathrm{MHz}$). This effect is enhanced
in the case of quasars because of the rapid rate of ionization and we see
that the entropy peaks around 150 $\mathrm{MHz}$ where the transition from a neutral
to an ionized Universe accelerates.

\section{Summary and Outlook}
\label{sec:conclusions}
We emphasized the need to expand the scenarios that need to be
explored for reionization and their role in building a reliable and
robust pipeline, thus necessitating fast realizations of different
scenarios.  We built a scheme in which a range
of 1-D ionization profiles were catalogued for a number of
luminosities, redshifts, densities and source spectra and which were
subsequently coupled to an $N$-body simulation to obtain an 
approximation of a ``standard'' \hbox{3-D} radiative transfer code. The
results obtained were validated using CRASH, a full \hbox{3-D} radiative
transfer code with ray tracing. The agreement between the two methods
was excellent for early redshifts ($z>8$), but as expected,
the discrepancy grew towards lower redshifts. Several
visual comparisons of the slices of different boxes were made along
with three different statistical measures of the similarity between
the simulations.

Many snapshots ($\approx 75$) were used between a redshift of 15 and 6
and the radiative transfer done on them as described in the preceding
sections. These snapshots were then used to make a contiguous cube
running from redshift 6 to 12. In terms of the observing frequency,
the cube spans 115 $\mathrm{MHz}$ to 200 $\mathrm{MHz}$ with a frequency resolution matching
that of LOFAR, about one $\mathrm{MHz}$. Cubes were generated for scenarios
involving only stars or quasars and some diagnostics are provided to
quantitatively differentiate between them. For both models the
variance in $\delta T_b$ peaked at around 160 $\mathrm{MHz}$. Although neither
may reflect reality, these scenarios demonstrate the use of the
techniques we have developed to span large parameter spaces of
variables. The PDF of $\delta T_b$ (see Fig.~\ref{fig:dtb_pdf})
provides a statistical discriminant between the different source
scenarios and could be used in the future to look for the statistical
detection of the signal.

The cubes generated provide $\delta T_b$ as a function of
frequency. The cube was then averaged over a bandwidth of 1~$\mathrm{MHz}$ and
convolved with the beam pattern of LOFAR to understand the distortions
caused by incomplete sampling by an interferometer. Even if the images
were blurred by the operation, the overall characteristics of the
signal remain detectable. Although the behaviour of the variance of
the signal before and after the convolution remained the same (except
that the latter had lower values on average), the image/slice entropy
showed a very different behaviour. In the former case, i.e., before the
convolution, the image entropy remained almost flat throughout the
frequency range whereas in the latter the entropy steeply rises at 
around 160 $\mathrm{MHz}$.

As a note, it is important to mention that although in principle
the ionized bubbles do move with a peculiar velocity $v_r(z) =
v_r(0)(1+z)^{-1/2}$, where $v_r(0)$ is the typical peculiar velocity
of galaxies at redshift zero, assuming $v_r(0) \approx 600$km/s, a
redshift 10 object would have a peculiar velocity of 200 km/s.
For a typical lifetime of the source considered, i.e., 10 Myr, this
corresponds to motion of about a couple of kpc. This is an order of
magnitude less than the resolution of the simulation box at that
redshift. Therefore we ignore this effect. On the other hand we have
taken into account the effect of redshift distortions whose effects
are relatively more important.

The simulations and comparisons in this paper have focused on purely
stellar or quasars sources, but it is plausible that the early sources
of reionization were a mixture of stars and quasars or other yet
unknown sources. It is therefore important to simulate reionization by
a mixture of these sources, taking into account their clustering
properties. The simulations presented in this paper did not take into
account the contraints on the population of ionizing sources imposed
by various measurements like the infrared excess in the case of stars
and the soft X-ray excess for quasars.  Apart from the ionization
patterns induced by these sources, the $\delta T_b$ maps will also
depend on the kinetic temperature which is coupled to the spin
temperature via collisions or Ly-$\alpha$ pumping. Hence, it is
imperative that we include these temperature effects on the IGM. We
will incorporate the mixture of sources and the effect of the
temperature in an upcoming paper (Thomas et al., \emph{in prep}).

One of the main astrophysical hurdles for the detection of the EoR signal
is the existence of prominent Galactic and extragalactic
foregrounds. Typically, the difference between the mean
amplitude of the EoR signal and the foregrounds is expected to be 4
to 5 orders of magnitude, but an interferometer like LOFAR measures
only the fluctuations which in this case are expected to be
different by `only' three orders of magnitude
\citep{jelic08}. In subsequent papers we will explore
the effects of foregrounds and their removal strategies.

\section{Acknowledgment}
LOFAR is being partially funded by the European Union, European Regional
Development Fund, and by ``Samenwerkingsverband Noord-Nederland'',
EZ/KOMPAS. The author RMT would also like to thank the Max Planck
Institute of Astrophysics, Garching, for the support provided during a
work visit to the institute.

\label{lastpage}


\begin{thebibliography}{99}
\bibitem[\protect\citeauthoryear{Bruzual \& Charlot}{2003}]{Bruzual:2003} 
 Bruzual G., Charlot S., 2003, MNRAS, 344, 1000
\bibitem[\protect\citeauthoryear{Ciardi, Ferrara, \& White}{2003}]{ciardi} 
 Ciardi B., Ferrara A., White S.~D.~M., 2003, MNRAS, 344, L7
\bibitem[\protect\citeauthoryear{Ciardi \& Madau}{2003}]{ciardi2} 
 Ciardi B., Madau P., 2003, ApJ, 596, 1
\bibitem[\protect\citeauthoryear{Ciardi et al.}{2001}]{crash} 
 Ciardi B., Ferrara A., Marri S., Raimondo G., 2001, MNRAS, 324, 381
\bibitem[\protect\citeauthoryear{Chabrier}{2003}]{Chabrier:2003} 
 Chabrier G., 2003, PASP, 115, 763 
\bibitem[\protect\citeauthoryear{Davis et al.}{1985}]{Davis:1985} 
 Davis M., Efstathiou G., Frenk C.~S., White S.~D.~M., 1985, ApJ, 292, 371
\bibitem[\protect\citeauthoryear{Dijkstra, Haiman, \& Loeb}{2004}]{dijkstra} 
 Dijkstra M., Haiman Z., Loeb A., 2004, ApJ, 613, 646 
\bibitem[\protect\citeauthoryear{Fan et al.}{2006}]{fan06}
 Fan X., et al., 2006, AJ, 131, 1203
\bibitem[\protect\citeauthoryear{Ferland et al.}{1998}]{Ferland:1998} 
 Ferland G.~J., Korista K.~T., Verner D.~A., Ferguson J.~W., Kingdon J.~B., 
 Verner E.~M., 1998, PASP, 110, 761 
\bibitem[\protect\citeauthoryear{Ferrarese}{2002}]{Ferrarese02} 
 Ferrarese L., 2002, ApJ, 578, 90
\bibitem[\protect\citeauthoryear{Fukugita \& Kawasaki}{1994}]{fuku94} 
 Fukugita M., Kawasaki M., 1994, MNRAS, 269, 563
\bibitem[\protect\citeauthoryear{Gnedin \& Abel}{2001}]{otvet} 
 Gnedin N.~Y., Abel T., 2001, NewA, 6, 437 
\bibitem[\protect\citeauthoryear{Hockney \& Eastwood}{1988}]{Hockney:1988} 
 Hockney R.~W., Eastwood J.~W., 1988, Computer Simulations Using Particles, 
 Taylor \& Francis
\bibitem[\protect\citeauthoryear{Pawlik \& Schaye}{2008}]{pawlik08} 
 Pawlik A.~H., Schaye J., 2008, arXiv, 802, arXiv:0802.1715 
\bibitem[\protect\citeauthoryear{Perley, Schwab,\& Bridle}{1989}]{perley89} 
 Perley R.~A., Schwab F.~R., Bridle A.~H., 1989, ASPC, 6,  
\bibitem[\protect\citeauthoryear{Hogan \& Rees}{1979}]{hogan79} 
 Hogan C.~J., Rees M.~J., 1979, MNRAS, 188, 791 
\bibitem[\protect\citeauthoryear{Jeli{\'c} et al.}{2008}]{jelic08} 
 Jeli{\'c} V., et al., 2008, MNRAS, 891
 \bibitem[\protect\citeauthoryear{Kuhlen \& Madau}{2005}]{kuhlen05} 
 Kuhlen M., Madau P., 2005, MNRAS, 363, 1069
\bibitem[\protect\citeauthoryear{Lambropoulous et al.,}{2008}]{panos08} 
 Lambropoulou P., et al.,  in prep
\bibitem[\protect\citeauthoryear{Madau, Meiksin, \& Rees}{1997}]{madau97} 
 Madau P., Meiksin A., Rees M.~J., 1997, ApJ, 475, 429
\bibitem[\protect\citeauthoryear{Maselli, Ferrara,\& Ciardi}{2003}]{maselli03} 
 Maselli A., Ferrara A., Ciardi B., 2003, MNRAS, 345, 379
\bibitem[\protect\citeauthoryear{Mellema et al.}{2006}]{c2ray} 
 Mellema G., Iliev I.~T., Alvarez M.~A., Shapiro P.~R., 2006, NewA, 11, 374
\bibitem[\protect\citeauthoryear{Mesinger \& Furlanetto}{2007}]{mesinger07} 
 Mesinger A., Furlanetto S., 2007, ApJ, 669, 663 
\bibitem[\protect\citeauthoryear{Nakamoto, Umemura, \& Susa}{2001}]{art} 
 Nakamoto T., Umemura M., Susa H., 2001, MNRAS, 321, 593
\bibitem[\protect\citeauthoryear{Nusser}{2005}]{nusser05} 
 Nusser A., 2005, MNRAS, 359, 183
\bibitem[\protect\citeauthoryear{Page et al.}{2007}]{page07} 
 Page L., et al., 2007, ApJS, 170, 335
\bibitem[\protect\citeauthoryear{Pritchard \& Furlanetto}{2007}]{pritchard07} 
 Pritchard J.~R., Furlanetto S.~R., 2007, MNRAS, 376, 1680
\bibitem[\protect\citeauthoryear{Razoumov \& Cardall}{2005}]{ftte} 
 Razoumov A.~O., Cardall C.~Y., 2005, MNRAS, 362, 1413
\bibitem[\protect\citeauthoryear{Rijkhorst et al.}{2006}]{flash} 
 Rijkhorst E.-J., Plewa T., Dubey A., Mellema G., 2006, A\&A, 452, 907
\bibitem[\protect\citeauthoryear{Ritzerveld, Icke, \& Rijkhorst}{2003}]{simpX} 
 Ritzerveld J., Icke V., Rijkhorst E.-J., 2003, astro, arXiv:astro-ph/0312301
\bibitem[\protect\citeauthoryear{Schaerer}{2002}]{schaerer} 
 Schaerer D., 2002, A\&A, 382, 28
\bibitem[\protect\citeauthoryear{Schaye \& Dalla Vecchia}{2008}]{Schaye:2008} 
 Schaye J., Dalla Vecchia C., 2008, MNRAS, 383, 1210
\bibitem[\protect\citeauthoryear{Scott \& Rees}{1990}]{scott90} 
 Scott D., Rees M.~J., 1990, MNRAS, 247, 510
\bibitem[\protect\citeauthoryear{Seljak \& Zaldarriaga}{1996}]{Seljak:1996} 
 Seljak U., Zaldarriaga M., 1996, ApJ, 469, 437
\bibitem[\protect\citeauthoryear{Spergel et al.}{2007}]{Spergel:2007}
 Spergel D.~N., et al., 2007, ApJS, 170, 377
\bibitem[\protect\citeauthoryear{Springel}{2005}]{Springel:2005} 
 Springel V., 2005, MNRAS, 364, 1105 
\bibitem[\protect\citeauthoryear{Springel \& Hernquist}{2003}]{springel03} 
 Springel V., Hernquist L., 2003, MNRAS, 339, 312 
\bibitem[\protect\citeauthoryear{Sunyaev \& Zeldovich}{1975}]{sz75} 
 Sunyaev R.~A., Zeldovich I.~B., 1975, MNRAS, 171, 375 
\bibitem[\protect\citeauthoryear{Susa}{2006}]{rsph} 
 Susa H., 2006, PASJ, 58, 445
\bibitem[\protect\citeauthoryear{Thomas \& Zaroubi}{2008}]{rajat08} 
 Thomas R.~M., Zaroubi S., 2008, MNRAS, 384, 1080 
\bibitem[\protect\citeauthoryear{Venkatesan, Giroux, \& Shull}{2001}]{venky01} 
 Venkatesan A., Giroux M.~L., Shull J.~M., 2001, ApJ, 563, 1
\bibitem[\protect\citeauthoryear{Whalen \& Norman}{2006}]{zeus} 
 Whalen D., Norman M.~L., 2006, ApJS, 162, 281
\bibitem[\protect\citeauthoryear{Wiersma, Schaye\& Smith}{2008}]{Wiersma:2008} 
 Wiersma R.~P.~C., Schaye J., Smith B.~D., 2008, MNRAS, submitted, 
 preprint (arXiv:0807.3748)
\bibitem[\protect\citeauthoryear{Zahn et al.}{2007}]{zahn07} 
 Zahn O., Lidz A., McQuinn M., Dutta S., Hernquist L., Zaldarriaga M., 
 Furlanetto S.~R., 2007, ApJ, 654, 12 
\bibitem[\protect\citeauthoryear{Zaroubi et al.}{2007}]{zaroubi07}
 Zaroubi S., Thomas R.~M., Sugiyama N., Silk J., 2007, MNRAS, 375, 1269 
\bibitem[\protect\citeauthoryear{Zaroubi \& Silk}{2005}]{zaroubi05} 
 Zaroubi S., Silk J., 2005, MNRAS, 360, L64
\end{thebibliography}
\end{document}